\documentclass{aa}
\newcommand{\xmm}{\textit{XMM-Newton}\xspace}
\newcommand{\chandra}{\textit{Chandra}\xspace}
\newcommand{\rosat}{\textit{ROSAT}\xspace}
\newcommand{\suzaku}{\textit{Suzaku}\xspace}
\usepackage{natbib,twoopt}
\usepackage{xcolor}
\usepackage[breaklinks=true, colorlinks=true, linkcolor=blue, citecolor=blue]{hyperref} %% to avoid \citeads line fills
\bibpunct{(}{)}{;}{a}{}{,}             %% natbib format for A&A and ApJ
\makeatletter
\newcommandtwoopt{\citeads}[3][][]{\href{http://adsabs.harvard.edu/abs/#3}%
    {\def\hyper@linkstart##1##2{}%
     \let\hyper@linkend\@empty\citealp[#1][#2]{#3}}}
  \newcommandtwoopt{\citepads}[3][][]{\href{http://adsabs.harvard.edu/abs/#3}%
    {\def\hyper@linkstart##1##2{}%
     \let\hyper@linkend\@empty\citep[#1][#2]{#3}}}
  \newcommandtwoopt{\citetads}[3][][]{\href{http://adsabs.harvard.edu/abs/#3}%
    {\def\hyper@linkstart##1##2{}%
     \let\hyper@linkend\@empty\citet[#1][#2]{#3}}}
  \newcommandtwoopt{\citeyearads}[3][][]%
    {\href{http://adsabs.harvard.edu/abs/#3}
    {\def\hyper@linkstart##1##2{}%
     \let\hyper@linkend\@empty\citeyear[#1][#2]{#3}}}
\makeatother
\usepackage{graphicx}
\usepackage{multirow}
\usepackage{comment}
\usepackage{siunitx}  %unit
\usepackage{txfonts}
\usepackage{subfig}
\usepackage{sidecap}
\sidecaptionvpos{figure}{c}
\begin{document}

   \title{The eROSITA view of the Abell 3391/95 field:}
   \subtitle{Cluster Outskirts and Filaments}

   \author{Angie Veronica
          \inst{1}
          \and
          Thomas H. Reiprich\inst{1}
          \and
          Florian Pacaud\inst{1}
          \and
          Naomi Ota\inst{1, 7}
          \and
          Jann Aschersleben\inst{1, 6}
          \and
          Veronica Biffi\inst{4, 5}
          \and
          Esra Bulbul\inst{2}
          \and
          Nicolas Clerc\inst{2}
          \and
          Klaus Dolag\inst{3}
          \and
          Thomas Erben\inst{1}
          \and
          Efrain Gatuzz\inst{2}
          \and
          Vittorio Ghirardini\inst{2}
          \and
          J\"urgen Kerp\inst{1}
          \and
          Matthias Klein\inst{2}
          \and          
          Ang Liu\inst{2}
          \and
          Teng Liu\inst{2}
          \and
          Konstantinos Migkas\inst{1, 8}
          \and
          Miriam E. Ramos-Ceja\inst{2}
          \and
          Jeremy Sanders\inst{2}
          \and
          Claudia Spinelli\inst{1}
          }

   \institute{Argelander-Institut f\"ur Astronomie (AIfA), Universit\"at Bonn, Auf dem H\"ugel 71, 53121 Bonn, Germany\\
              \email{averonica@astro.uni-bonn.de}
        \and
        Max-Planck-Institut f\"ur extraterrestrische Physik, Gießenbachstraße 1, 85748 Garching, Germany
        \and
        Universit\"ats-Sternwarte, Fakult\"at f\"ur Physik, Ludwig-Maximilians-Universit\"at M\"unchen, Scheinerstr.1, 81679 München, Germany 
        \and
        INAF - Osservatorio Astronomico di Trieste, via Tiepolo 11, I-34143 Trieste, Italy
        \and
        IFPU - Institute for Fundamental Physics of the Universe, Via Beirut 2, I-34014 Trieste, Italy
        \and
        Kapteyn Astronomical Institute, University of Groningen, PO Box 800, 9700 AV Groningen, The Netherlands
        \and
        Nara Women's University, Kitauoyanishi-machi, Nara, 630-8506, Japan
        \and
        Leiden Observatory, Leiden University, PO Box 9513, NL-2300 RA Leiden, The Netherlands}
   \date{Received ; accepted}

% \abstract{}{}{}{}{} 
% 5 {} token are mandatory
 
  \abstract
   {About $30\%-40\%$ of the baryons in the local Universe remain unobserved. Many of these "missing" baryons are expected to reside in the warm-hot intergalactic medium (WHIM) of the cosmic web filaments that connect clusters of galaxies. SRG/eROSITA performance verification (PV) observations covered about 15 square degrees of the A3391/95 system and have revealed a $\sim$15 Mpc continuous soft emission connecting several galaxy groups and clusters.}
   {We aim to characterize the gas properties in the cluster outskirts ($R_{500}<r<R_{200}$) and in the detected inter-cluster filaments ($>R_{200}$) and to compare them to predictions.}
   {We performed X-ray image and spectral analyses using the eROSITA PV data in order to assess the gas morphology and properties in the outskirts and the filaments in the directions of the previously detected Northern and Southern Filament of the A3391/95 system. We constructed surface brightness profiles using particle-induced background-subtracted, exposure- and Galactic absorption-corrected eROSITA products in the soft band ($0.3-2.0~\mathrm{keV}$). We constrained the temperatures, metallicities, and electron densities through X-ray spectral fitting and compared them with the expected properties of the WHIM. We took particular care of the foreground.}
   {In the filament-facing outskirts of A3391 and the Northern Clump, we find higher temperatures than typical cluster outskirts profiles, with a significance of between $1.6-2.8\sigma$, suggesting heating due to their connections with the filaments. We confirm surface brightness excess in the profiles of the Northern, Eastern, and Southern Filaments. From spectral analysis, we detect hot gas of $0.96_{-0.14}^{+0.17}~\mathrm{keV}$ and $1.09_{-0.06}^{+0.09}~\mathrm{keV}$ for the Northern and Southern Filament, respectively, which are close to the upper WHIM temperature limit. The filament metallicities are below 10\% solar metallicity and the electron densities are ranging between $2.6$ and $6.3\times10^{-5}~\mathrm{cm^{-3}}$. The characteristic properties of the Little Southern Clump (LSC), which is located at a distance of $\sim\!1.5R_{200}$ from A3395S in the Southern Filament, suggest that it is a small galaxy group. Excluding the LSC from the analysis of the Southern Filament does not significantly change the temperature or metallicity of the gas, but it decreases the gas density by 30\%. This shows the importance of taking into account any clumps in order to avoid overestimation of the gas measurement in the outskirts and filament regions.}
   {We present measurements of morphology, temperature, metallicity, and density of individual warm-hot filaments. The electron densities of the filaments are consistent with the WHIM properties as predicted by cosmological simulations, but the temperatures are higher. As both filaments are short (1.8 and 2.7 Mpc) and located in a denser environment, stronger gravitational heating may be responsible for this temperature enhancement. The metallicities are low, but still within the expected range from the simulations.}


   \maketitle
%
%-------------------------------------------------------------------

\section{Introduction}
\begin{SCfigure*}[][!h]
%\centering
\includegraphics[width=0.75\textwidth]{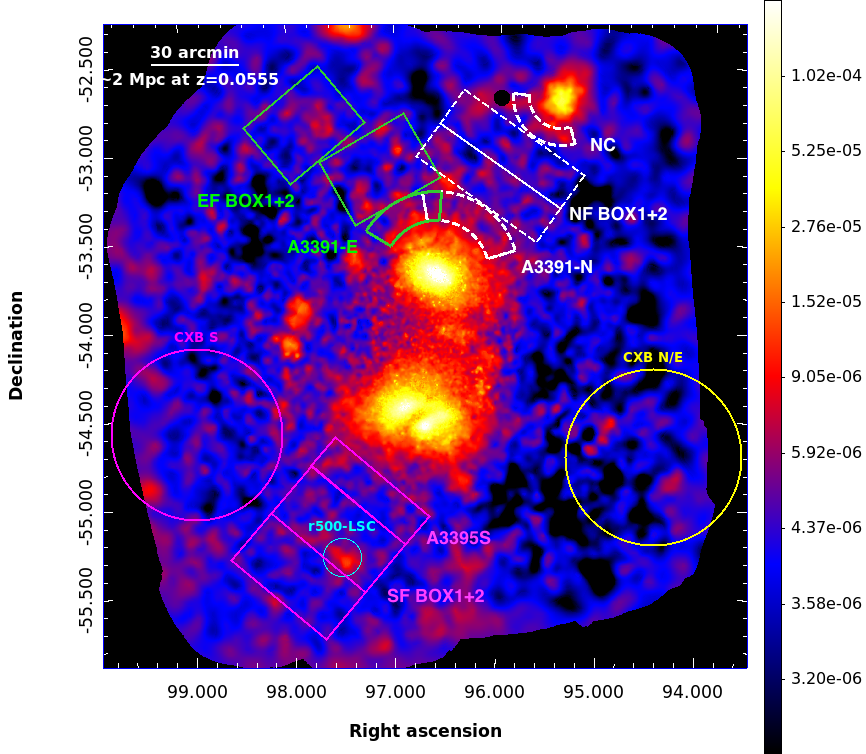}
\hspace{5pt}
\caption{eROSITA PIB-subtracted, exposure- and Galactic-absorption-corrected image in the $0.3-2.0~\mathrm{keV}$ band. The image has been adaptively smoothed with S/N set to 12. The point sources have been removed and refilled with their surrounding background values. The defined regions are used for spectral analyses (see Sec. \ref{sec:spectro}). The colorbar is in the units of counts per second (as well as the following X-ray images).}
\label{fig:spectro_regions}
\end{SCfigure*}

The missing baryon problem states that about $30\%-40\%$ of the total baryons in the Universe are still unaccounted for \citep[e.g.,][]{Shull_2012}. According to the large-scale cosmological simulations \citep[e.g.,][]{Cen_1999, Dave_2001}, they should be found in the warm-hot intergalactic medium (WHIM) located in the so-called cosmic web filaments \citep{Bond_1996}. These filaments connect galaxy clusters and play an important role in transferring matter onto the clusters, thus supporting their growth \citep{West_1995, Tanaka_2007, Bond_2010-filament}.
\par
Due to the faint nature of the WHIM emission (the electron densities and temperature ranging between $n_\mathrm{e}\approx10^{-6}-10^{-4}~\mathrm{ cm}^{-3}$ and $T = 10^5-10^7$ K \citep{Nicastro_2017}, respectively) detecting the WHIM is challenging. However, with the X-ray emissivity being proportional to the square of the density, it is reasonable to direct attempts at the densest part of this tenuous gas, for example between a pair of galaxy clusters (e.g., \citealt{Kull_1999, Yutaka_2008, Vazza_2019}), as well as at the outskirts of a galaxy cluster \citep[e.g.,][]{Reiprich_2013, Nicastro_2018}. Studies of the WHIM at these locations include measurements of the filament between the Abell 222 and 223 cluster pair \citep{Werner_2008}, the Abell 399 and 401 system \citep[e.g.,][]{Akamatsu_2017, Bonjean_2018}, the Abell 98N/S system \citep{Arnab_2022, Alvarez_2022}, the outskirts of the Coma cluster \citep{Bonamente_2009, Bonamente_2022}, the Abell 2744 cluster \citep{Eckert_2015, Hattori_2017}, and the Abell 1750 cluster \citep{Bulbul_2016}.
\par
The binary galaxy cluster system Abell 3391 and Abell 3395 (A3391/95) is also one of the systems where a search for the WHIM has been carried out, and the system has therefore been observed extensively using various instruments of different wavelengths, including several X-ray telescopes;for example by \rosat and \textit{ASCA} and more recently by \xmm, \chandra, and \suzaku. Earlier studies showed that the gas between the A3391 and the double-peaked A3395S/N (hereafter referred to as the bridge) is consistent with a filament \citep{Tittley_2001}. While latter studies concluded that the gas properties in these regions are more typical of tidally stripped intracluster medium (ICM) gas from A3391 and A3395, a sign of an early cluster merger phase \citep{Sugawara_2017, Alvarez_2018}. A study of the A3391/95 system was also carried out using \textit{Planck} data, and a significant excess of signal near the A3395S/N system in the thermal Sunyaev-Zel'dovich (tSZ) residual map was reported \citep{Planck_2013}. However, the individual components of this cluster cannot be resolved by the data due to the modest resolution of Planck.
\par
The extended ROentgen Survey with an Imaging Telescope Array \citep[eROSITA,][]{Predehl_2021} is the newly launched German X-ray telescope on board the Spectrum-Roentgen-Gamma (SRG) mission. It has a wide field-of-view (FoV) of 1.03 degrees in diameter. In the soft energy band of $0.5-2.0~\mathrm{keV}$, the effective area of the seven combined eROSITA telescope modules (TMs) is slightly higher than the combined \xmm cameras (MOS1+MOS2+pn) \citep{Predehl_2021}. Additionally, eROSITA operates in the scan and survey modes, which allows us to easily reach the virial radii of clusters and to constrain fore- and background emission. The FoV-averaged PSF ($\sim\!26''$) is significantly better than that of the \textit{Suzaku} satellite, allowing us to remove unrelated background sources to much lower flux levels. The energy resolution at soft energies is better than for any other operating X-ray satellite CCD, allowing us to better characterize soft band emission lines. And the short focal length of 1.6m minimizes the particle-induced background (PIB) levels. These properties make eROSITA an excellent instrument for studying the outskirts of galaxy clusters and finding large-scale structures.
\par
In October 2019, the A3391/95 system was observed as one of the eROSITA performance verification (PV) targets. The field was observed four times during this phase, with three raster-scan observations and one pointing observation. The combination of all four observations covers an area of about $15$ square degrees, which gives us the deepest large-scale X-ray view of the system to date. Using these observations, eROSITA revealed indications of the presence of warm gas in the bridge region, on top of the tidally stripped hot ICM gas \citep{Reiprich_2021}. This finding is also seen in the analogous A3391/95 cluster system from Magneticum simulations \citep{Biffi_2022}. Furthermore, a continuous soft emission extending from the north to the south was discovered. This $\sim$15 Mpc filamentary structure seems to connect at least five galaxy groups and clusters.
Those are A3391, A3395S, A3395N, the Northern Clump\footnote{The extended objects detected in the A3391/95 field are referred to as "clumps", although, the Northern Clump, for example, is a galaxy cluster.} \citep[NC; MCXC J0621.7-5242 cluster or MS 0620.6-5239 cluster,][]{Tittley_2001, Piffaretti_2011, Veronica_2022}, the Little Southern Clump (LSC), and the MCXC J0631.3-5610 cluster. The LSC is embedded in the Southern Filament and has fully entered the $2R_{200}$ of A3395, while the center of the MCXC J0631.3-5610 cluster is just outside the FoV of these PV observations.
\par
In this work, we utilize the eROSITA A3391/95 PV data to investigate the gas properties of the outskirts and the detected filaments. These filaments are the Northern and Eastern Filaments extending north and northeastward from the A3391 cluster, and the Southern Filament in the south of the A3395S/N cluster. The present paper is structured as follows: in Sect.~\ref{sec2}, we describe our observations, the data reduction steps, and our analysis strategy. In Sect.~\ref{sec3}, we present the results of our imaging analysis (Sect.~\ref{sec3:surb}) and those of our spectral analysis (Sect.~\ref{sec3:spectro}). In Sect.~\ref{sec:discuss}, we discuss these results. Finally, in Sect.~\ref{sec:summarize}, we summarize our findings and conclude the results.
\par
Unless stated otherwise, all uncertainties are at the 68.3\% confidence interval. The assumed cosmology in this work is a flat $\Lambda$CDM cosmology adopted from \cite{Planck2018_2020}, such that the Hubble constant $H_0=67.4~\mathrm{km~s^{-1}~Mpc^{-1}}$, $\Omega_\mathrm{m}=0.315$, $\Omega_\Lambda = 0.6847$, and $\Omega_\mathrm{b}=0.0493$. At the redshift of A3391 cluster, $z=0.0555$, $1''$ corresponds to 1.119 kpc.

%--------------------------------------------------------------------
\section{Data reduction and analysis}\label{sec2}
\begin{table}
    \centering
    \caption{eROSITA A3391/95 PV observations used in this work. There are 16 datasets in total among the observations.}
    \resizebox{\columnwidth}{!}{\begin{tabular}{c c c c}
    \hline
    \hline
Observing Date & ObsID (mode) & TM & Exposure$^*$ [ks] \\
\hline
October 2019 & 300005 (scan) & 5-7 & 55 \\
October 2019 & 300006 (scan) & 5-7 & 54 \\
October 2019 & 300016 (scan) & 1-7 & 58 \\
October 2019 & 300014 (pointed) & 5-7 & 35\\
%8-9 March2020
\hline
\multicolumn{4}{l}{\footnotesize $^*$The exposure times listed are the average filtered on-axis exposure}\\
\multicolumn{4}{l}{\footnotesize time across the available TMs of each observation.}\\
\hline
\hline
    \end{tabular}}
    \label{tab:obs}
\end{table}

\begin{table*}
    \centering
    \caption{Information of the A3391/95 groups and clusters.}
    \begin{tabular}{c c c c c c}
    \hline
    \hline
\multirow{3}{*}{Object} & \multirow{2}{*}{R.A.} & \multirow{2}{*}{Dec.} & \multirow{3}{*}{$z$} & $R_{500}$ & $R_{200}$ \\
 & & & & $[']$ & $[']$ \\ 
 & (J2000) & (J2000) & & [Mpc] & [Mpc] \\
\hline \\[-1.7ex]
\multirow{2}{*}{A3391$^a$} & \multirow{2}{*}{06h 26m 20.86s} & \multirow{2}{*}{-53d 41m 30.48s} & \multirow{2}{*}{0.0555} & 18.090 & 27.830\\
 & & & & 1.170 & 1.800 \\
\hline
\multirow{2}{*}{A3395N$^a$} & \multirow{2}{*}{06h 27m 37.56s} & \multirow{2}{*}{-54d 26m 48.12s} & \multirow{2}{*}{0.0518} & 22.81 & 35.09\\
 & & & & 1.383 & 2.128 \\
\hline
\multirow{2}{*}{A3395S$^a$} & \multirow{2}{*}{06h 26m 48.58s} & \multirow{2}{*}{-54d 32m 45.60s} & \multirow{2}{*}{0.0517} & 22.94 & 35.29\\
 & & & & 1.388 & 2.136 \\
\hline
Northern$^b$ & \multirow{2}{*}{06h 21m 43.34s} & \multirow{2}{*}{-52d 41m 33.00s} & \multirow{2}{*}{0.0511} & 10.620 & 16.340\\
Clump & & & & 0.636 & 0.978 \\
\hline
Little$^c$ & \multirow{2}{*}{06h 30m 04.80s} & \multirow{2}{*}{-55d 17m 51.50s} & \multirow{2}{*}{0.0562} & 6.44 & 10.03\\
Southern Clump & & & & 0.437 & 0.682 \\
    \hline
\hline
\multicolumn{6}{l}{\footnotesize $^a$\cite{Reiprich_2021}, $^b$\cite{Veronica_2022}, $^c$this work}\\
    \hline
    \hline
    \end{tabular}
    \label{tab:clusters}
\end{table*}

All eROSITA A3391/95 PV observations are listed in Table \ref{tab:obs}. We used eROSITA data processing in configuration c001. The data reduction steps of all 16 eROSITA data sets were performed using the extended Science Analysis Software (eSASS, \citealt{Brunner_2022}) version \texttt{eSASSusers\_201009}. The data reduction steps and the image-correction steps, were done following the procedure described in Sect. 2.1 and 3.3 of \cite{Reiprich_2021}, which include the particle-induced background (PIB) subtraction, exposure correction (including vignetting), and Galactic absorption correction across the FoV.

\subsection{Imaging analysis}\label{sec:imag_analysis}
To focus on the soft emission in the outskirts and from the filaments, we used the energy band of $0.3-2.0~\mathrm{keV}$ for the imaging analysis. The lower energy limit used for the telescope modules (TMs) with an on-chip filter (TM1, 2, 3, 4, 6; the combination of these TMs is referred to as TM8) was set to 0.3 keV, while for the TMs without an on-chip filter (TM5 and 7; the combination of these TMs is referred to as TM9) due to the optical light leak contamination \citep{Predehl_2021}, the lower energy limit was set to 0.5 keV. The count rates of the final image (all observations combined, and fully corrected) correspond to an effective area given by one TM with an on-chip filter in the energy band $0.3-2.0~\mathrm{keV}$.
\par
The Galactic absorption across the FoV was corrected by calculating the ratio of the expected eROSITA count rates with the total hydrogen column density ($N_\mathrm{H}$) at a given position to the count rate of the median $N_\mathrm{H}$ across the FoV. The count rates of different $N_\mathrm{H}$ values were estimated assuming a typical X-ray fore- and background model, that is one unabsorbed with one absorbed from diffuse thermal components and one power law component. These components represent the Local Hot Bubble (LHB), the Milky Way halo (MWH), and unresolved sources, respectively. The computed $N_\mathrm{H}$ values are from the A3391/95 total $N_\mathrm{H}$ map \citep[for more details, see Sect. 2.5 of][]{Reiprich_2021} generated from the IRAS $100~\mathrm{\mu m}$ \citep{IRAS_2005} and the HI4PI data \citep{HI4PI_2016}. A ratio map relative to the median $N_\mathrm{H}$ was generated for each type of eROSITA TMs. We divided the exposure map by their corresponding ratio map for the Galactic absorption correction.
\par
The point source catalog was generated using the method described in \cite{Pacaud_2006} and \cite{Ramos-Ceja_2019}, that is, the point sources were detected on a wavelet-filtered image by the Source Extractor software \citep{SExtractor}. Using the obtained catalog, we excluded them from our analyses. The so-called ghosting procedure was also performed to generate the surface brightness image. Through this procedure, the areas from which the point sources were excised were refilled with surrounding background photons. Any ghosted images generated in this work were only used for visualization purposes. To avoid any biases introduced by using the artificially refilled pixels in the analysis, we excluded these pixels from any calculations and the missing area was properly taken into account. The final PIB-subtracted, point-sources-subtracted and refilled, exposure-corrected, and Galactic-absorption-corrected image is shown in Fig.~\ref{fig:spectro_regions}. The image has been adaptively smoothed with a signal-to-noise ratio (S/N) set to 12 in order to enhance low-surface-brightness emission and large-scale features.
\par
The X-ray surface brightness profiles were calculated using the final eROSITA products in the $0.3-2.0~\mathrm{keV}$ energy band, including the photon image, the fully corrected exposure map, and the PIB map, as well as the point source catalog. The large eROSITA FoV and the SRG scan mode in which most of the observations were taken, allow us to model the cosmic X-ray background (CXB) from regions within the FoV. To account for the variation of the CXB, we placed ten boxes across the field. To ensure minimum emission from the clusters and the filaments, which could cause overestimation in the CXB level (and therefore underestimation of the surface brightness values), they were placed beyond the $R_{200}$\footnote{All the cluster radii in this work are calculated using the relations stated in \cite{Reiprich_2013}, e.g., $r_{500}\approx0.65r_{200}$, $r_{100}\approx1.36r_{200}$, and $r_{2500}\approx0.28r_{200}$.} of each cluster and the defined area of the filaments. The configuration of these CXB boxes can be found in Fig. F.1. of \cite{Reiprich_2021}. The average of the surface brightness values of these ten boxes was taken to be the CXB level in the field, $\mathrm{SB}_{\mathrm{CXB}}$, while the root-mean-square deviation of these values was taken as the standard deviation, $\sigma_{\mathrm{CXB}}$, which is larger than the statistical uncertainty. We report that the CXB surface brightness value in the A3391/95 system is of the typical value. The CXB value we obtained is of the same order of magnitude as, for example, to the CXB surface brightness value calculated for the A3158 surface brightness profile using the eROSITA calibration data \citep{Whelan_2022}, which is not affected by additional foreground components, such as the eROSITA Bubble or other extended Galactic sources.
\par
In their review, \cite{Reiprich_2013} define the outskirts of galaxy clusters as the regions found within the range of $R_{500}<r<3R_{200}$. In the present work, we also define the lower boundary of the outskirts region to be $R_{500}$, but as the $3R_{200}$ of the A3391/95 clusters extends to the entire FoV of the eROSITA PV data, we chose an arbitrary upper boundary of $R_{200}$. The filament was defined as the inter-cluster emission beyond the $R_{200}$ of a cluster. The used cluster radii, as well as their corresponding emission peak coordinates and their references, are listed in Table \ref{tab:clusters}. The $R_{500}$ of A3391 and A3395N/S were calculated from \rosat and \textit{ASCA} observations \citep{Reiprich_2002}; the $R_{500}$ of the NC was estimated from a dedicated \xmm observation \citep{Veronica_2022}; and the $R_{500}$ of the LSC is a product of the present study (eROSITA PV data; see Sec. \ref{Sec:LSC}).

\subsection{Spectral analysis}\label{sec:spectro}
We utilized the eSASS task \texttt{srctool} to extract all eROSITA spectra, the Ancillary Response Files (ARFs), and the Response Matrix Files (RMFs) for the source and background regions. The spectral fitting was performed with \texttt{XSPEC} \citep{XSPEC} version: 12.10.1f.
\par
For the spectral analysis, we used the third eROSITA scan observation (ObsID: 300016), where all seven eROSITA TMs are available. The configuration of the regions used in the spectral analysis is shown in Fig.~\ref{fig:spectro_regions}. Any source regions related to the Northern (Eastern) Filament are indicated with dashed white (green) color. For instance, the A3391 outskirts region ($R_{500}-R_{200}$) in the direction of the Northern (Eastern) Filament is plotted as a white (green) sector, labeled A3391-N (A3391-E). The Northern (Eastern) Filament source regions are represented by two white (green) boxes, labeled NF BOX1+2 (EF BOX1+2). In the south, the source regions are depicted with magenta boxes, labeled A3395S and SF BOX1+2 for the outskirts and filament regions, respectively. The cyan circle marks the $R_{500}$ of the LSC. For each filament, we placed two identical boxes in line with one another outside the $R_{200}$ of the corresponding parent clusters. For the analyses of the Northern and the Eastern Filament, we took the average redshift between the A3391 cluster and the NC, $z_{N/E} = 0.0533$. While for the Southern Filament, we used the average redshift between the A3395N, A3395S, and the MCXC J0631.3-5610 clusters, $z_S = 0.0525$. The height ($h$) and width ($2r$) of each box are $14'~(0.90~\mathrm{Mpc})$ and $50'~(3.23~\mathrm{Mpc})$ for the Northern Filament, $25'~(1.62~\mathrm{Mpc})$ and $33'~(2.13~\mathrm{Mpc})$ for the Eastern Filament, and $20.83'~(1.33~\mathrm{Mpc})$ and $41.67'~(2.66~\mathrm{Mpc})$ for the Southern Filament, respectively. 
\par
The X-ray spectral fitting model, including the CXB and the source emission, is described in the following equation:
\begin{equation}
\begin{split}
\mathtt{Model =} &\quad\mathtt{(apec_1 + TBabs\times(apec_2 +}\\
&\quad\mathtt{powerlaw)) + TBabs\times apec_3} \mathtt{~+~PIB},
\end{split}
\label{eq:spectral_model}
\end{equation}
where the first term represents the CXB components. The absorption parametrized by the hydrogen column density along the line of sight, $N_\mathrm{H}$, is represented by \texttt{TBabs} \citep{Wilms_2000}. The $N_\mathrm{H}$ values used in this work are taken from the A3391/95 total $N_\mathrm{H}$ map (see Sect.~\ref{sec:imag_analysis}). The X-ray foreground emission from the LHB and the MWH are described by \texttt{apec$\mathtt{_1}$} and \texttt{apec$\mathtt{_2}$}, respectively. Lastly, the cosmic X-ray background from the unresolved sources \citep[e.g.,][]{Luo_2017} is represented by the power law component, \texttt{powerlaw}.
\begin{table}
    \centering
    \caption{Information on the parameters of the eROSITA CXB components and their best-fit normalization values from the first fitting step (see Sect.~\ref{sec:spectro}).}
    \resizebox{\columnwidth}{!}{\begin{tabular}{c c c }
    \hline
    \hline
Component & Parameter & Value\\
    \hline
\texttt{TBabs} &  $N_\mathrm{H}^{\dagger, N/E}$ & 0.046\\
& $N_\mathrm{H}^{\dagger, S}$  & 0.077\\
\hline
\texttt{apec$\mathtt{_1}$} (LHB) & $k_\mathrm{B}T$ [keV] & 0.1\\
& $Z~[Z_\odot]$ & 1\\
& $z$ & 0\\
& $norm^{*, N/E}$ & $(1.48\pm0.12)\times10^{-6}$\\[5pt]
& $norm^{*, S}$ & $(1.80_{-0.10}^{+0.08})\times10^{-6}$\\[5pt]
\hline
\texttt{apec$\mathtt{_2}$} (MWH) &  $k_\mathrm{B}T$ [keV] & 0.25\\
& $Z~[Z_\odot]$ & 1 \\
& $z$ & 0\\
& $norm^{*, N/E}$ & $(3.63_{-0.28}^{+0.25})\times10^{-7}$\\[5pt]
& $norm^{*, S}$ & $(5.90_{-0.25}^{+0.23})\times10^{-7}$\\[5pt]
\hline
\texttt{powerlaw} & $\Gamma$ & 1.46\\
(unresolved sources) & $norm^{\ddagger, N/E}$ & $(5.82_{-0.18}^{+0.20})\times10^{-7}$\\[5pt]
& $norm^{\ddagger, S}$ & $(5.63_{-0.15}^{+0.16})\times10^{-7}$\\[5pt]
\hline
\multicolumn{3}{l}{\footnotesize $^{N/E}$ from CXB used for the Northern or Eastern Filament analysis,}\\
\multicolumn{3}{l}{\footnotesize $^{S}$ from CXB used for the Southern Filament analysis,}\\
\multicolumn{3}{l}{\footnotesize $^\dagger$[10$^{22}$ atoms cm$^{-2}$], $^*[\mathrm{cm}^{-5}/\mathrm{arcmin}^2]$, $^{\ddagger}$[photons/keV/cm$^2$/s/arcmin$^2$ at 1 keV]}\\
    \hline
    \hline
    \end{tabular}}
    \label{tab:sky_BG}
\end{table}

\begin{figure}
\centering
\includegraphics[width=\columnwidth]{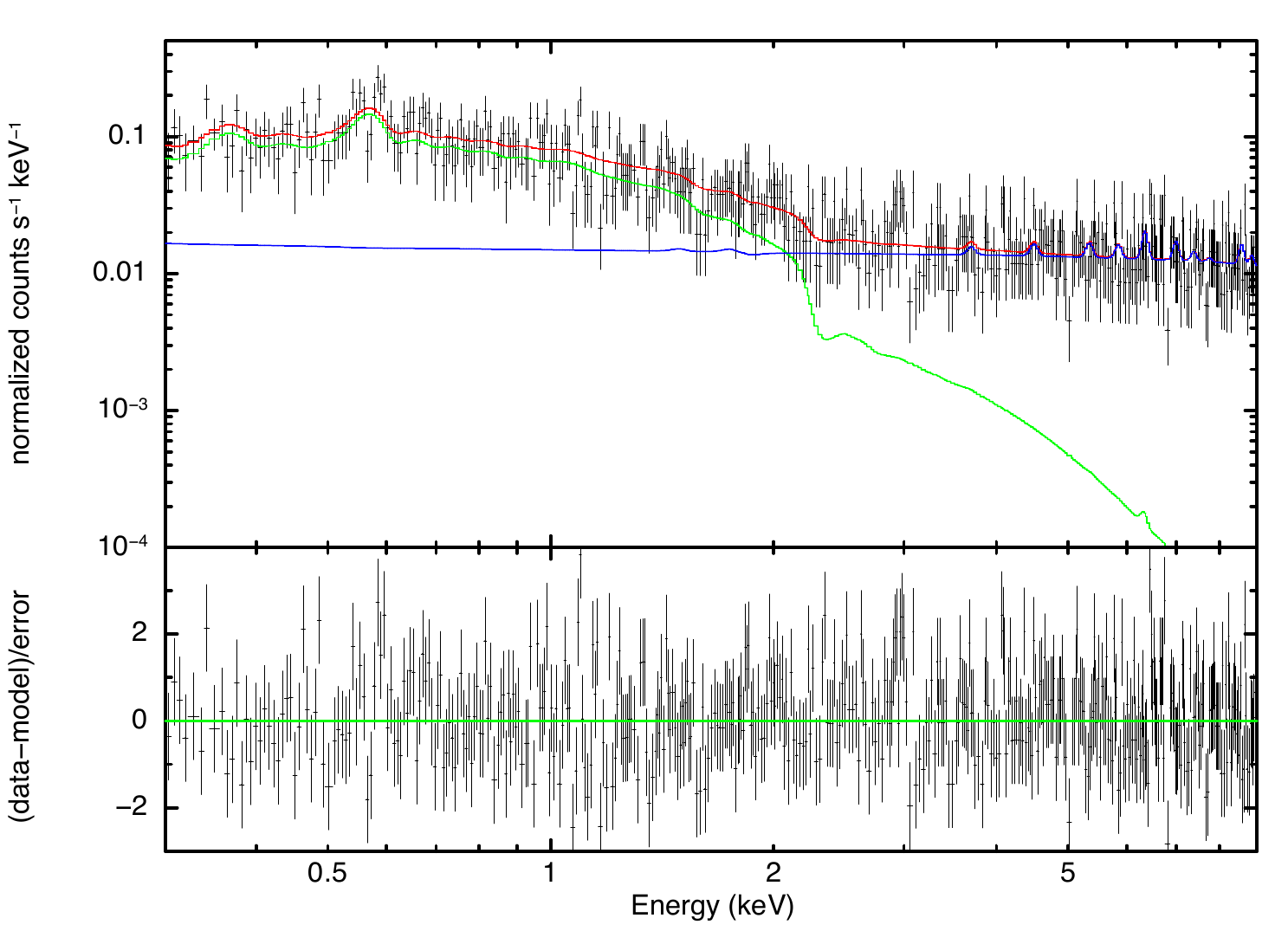}
\caption{eROSITA spectrum of the northern outskirts sector of A3391 (A3391-N). The spectra and the corresponding response files of TM3 and TM4 are merged for better visualization. The black points are the spectral data, while the red, green, and blue lines represent the total model, the cluster and CXB model, and the instrumental background model, respectively.}
\label{fig:eROSITA_spectrum}
\end{figure}

The second term, \texttt{TBabs}$\times$\texttt{apec$\mathtt{_3}$}, represents the absorbed cluster or filament emission. The \texttt{apec} parameters include temperature $k_\mathrm{B}T$, metallicity $Z$, redshift $z$, and normalization $norm$. The parameter $norm$ is defined as

\begin{equation}
norm = \frac{10^{-14}}{4\pi[D_A(1+z)]^2}\int n_\mathrm{e} n_\mathrm{H} \mathrm{d}V,
\label{eq:norm}
\end{equation}
where $D_A$ is the angular diameter distance to the source in centimeters (cm), while $n_\mathrm{e}$ and $n_\mathrm{H}$ are the electron and hydrogen densities in cm$^{-3}$, respectively.
\par
To take into account the $N_\mathrm{H}$ variations in the FoV, two different CXB regions were used. These were defined based on how closely the average $N_\mathrm{H}$ (see Sect.~\ref{sec:imag_analysis}) values within them resemble the average $N_\mathrm{H}$ values of the sources. The first region (Fig.~\ref{fig:spectro_regions}, yellow circle) was used for the Northern and Eastern Filament, and the second (magenta circle) was used for the Southern Filament. We fixed the $N_\mathrm{H}$ values of the source and CXB spectra to their respective average values in the X-ray spectral fitting. The placement of the source and CXB regions on the $N_\mathrm{H}$ map is shown in Fig.~\ref{fig:NH} of Appendix~\ref{App:A}. In this figure, we also labeled the used average $N_\mathrm{H}$ value of each region.
\par
The third term of Eq.~\ref{eq:spectral_model}, \texttt{PIB}, represents the instrumental background model. We utilized the results of the eROSITA EDR Filter Wheel Closed (FWC)\footnote{\href{https://erosita.mpe.mpg.de/edr/eROSITAObservations/EDRFWC/}{https://erosita.mpe.mpg.de/edr/eROSITAObservations/EDRFWC/}} data analysis to model the instrumental background of TM8 and the modified version for TM9, where the parameters were constrained including the lower-energy FWC spectra. The components of the instrumental background consist of a combination of a power law and an exponential cut-off to model the signal above $\sim$1 keV and two power laws to model the background increase due to the detector noise at low energy. Additionally, 14 Gaussian lines are included to model the fluorescence lines caused by the interaction between the cosmic particles with the detector components. Among these, the brightest lines are Al-K$_\alpha$ at 1.486 keV and Fe-K$_\alpha$ at 6.391 keV, and the weaker lines are Ti-K$_\alpha$ at 4.504 keV, Co-K$_\alpha$ at 6.915 keV, Ni-K$_\alpha$ at 7.461, Cu-K$_\alpha$ at 8.027 keV, and Zn-K$_\alpha$ at 8.615 keV. Further details of the eROSITA in-flight background are discussed in \cite{Freyberg_2022}.
\par
We performed the fitting in the energy band of $0.3-9.0~\mathrm{keV}$ for the TMs with an on-chip filter. Below 0.3 keV, the detector noise becomes stronger, while no strong source signal is expected above 9.0 keV given eROSITA's effective area. We find that the light leak contamination in the A3391/95 eROSITA PV observations is lower than in the other Cal/PV observations. Based on the light-leak diagnostic of this field, we can use a low energy limit of 0.5 keV for TM5 and TM7.
\par
The spectral fitting procedure started with fitting the CXB spectra of all TMs in order to obtain best-fit values and the errors on the normalizations of the CXB components. The information from this first fitting in the respective CXB regions, including the CXB components and the resulting best-fit values are listed in Table \ref{tab:sky_BG}. Subsequently, the source and CXB spectra of all TMs were fitted simultaneously. In this second fitting, we freed the \texttt{apec$\mathtt{_3}$} parameters, $k_\mathrm{B}T$, $Z$, and $norm$, of the source spectra whenever the statistics allowed. Otherwise, we fixed $Z$ to 0.3 of \cite{Asplund_2009} for the cluster outskirts (\citealp[e.g.,][]{Reiprich_2013, Urban_2017}), and 0.1 and 0.2 for the filaments (\citealp[e.g.,][]{Tanimura_2020, Biffi_2022}). The normalizations of the instrumental background components were also freed. The normalizations of the CXB components in this second fitting were also initially freed. However, we found one case where the fit was seen to be driven by one of the CXB components, that is the Eastern Filament fit with metallicity fixed to 0.1 solar. In this case, the obtained temperature for the Eastern Filament is 0.23 keV, which is close to the MWH temperature of 0.25 keV. This temperature drop is accompanied by an increase in source normalization, while the MWH normalization decreases. Translating the Eastern Filament normalization value in this case to the gas overdensity ($\delta_b$), we find $\delta_b\approx356$ (3.5 times higher than the value obtained when the CXB normalizations are restricted; see Table~\ref{tab:spectro}). This value is not physical for cluster outskirts nor filament regions, but more common for the hot and denser ICM gas. The MWH normalization in this case drops by $18\%$. Therefore, to avoid any source parameters being driven by the CXB components, we restricted the CXB normalizations of the source spectra to the minimum and maximum values obtained from the CXB spectra from the first fit. We note that the resulting gas properties of the outskirts and filaments between freeing and restricting the CXB normalizations are consistent within the $1\sigma$ uncertainties, except for the Eastern Filament case described above. In this work, we present only the results from the restricted source CXB normalization fit. The C-statistic \citep{Cash_1979} was adopted. An example of the eROSITA spectrum and its fitted model are shown in Fig.~\ref{fig:eROSITA_spectrum}.
%%%%%%%%%%%%%%%%%%%%%%%%%%%%%%%%%%%%%%%%%%%%%%%%%%%%%%%%
%%%%%%%%%%%%%%%%%%%%%%%%%%%%%%%%%%%%%%%%%%%%%%%%%%%%%%%%
\section{Results}\label{sec3}
\subsection{X-ray surface brightness profiles}\label{sec3:surb}
\subsubsection{Configuration}
\begin{figure*}[h!]
\centering
\includegraphics[width=0.48\textwidth]{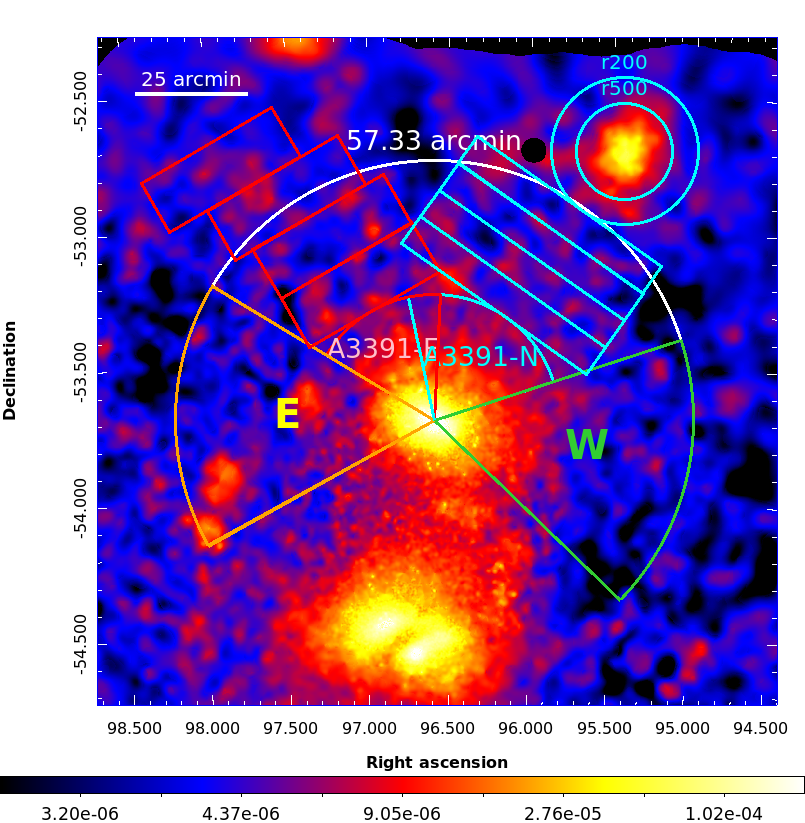}
\includegraphics[width=0.48\textwidth]{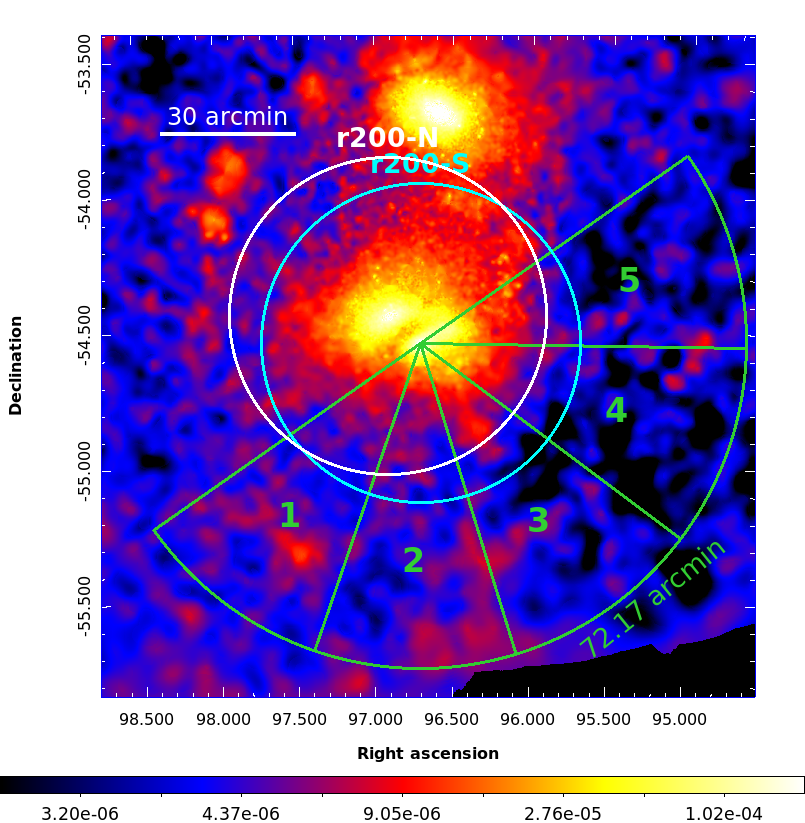}
\caption{Configuration of surface brightness profiles related to the A3391 cluster, NC, Northern and Eastern Filament (\textit{left}), and the A3395 cluster and Southern Filament (\textit{right}) overlaid on the adaptively-smoothed eROSITA PIB subtracted, exposure corrected, and Galactic absorption corrected image in the $0.3-2.0$ keV band.}
\label{fig:SBconfig}
\end{figure*}

\begin{figure*}
\centering
\subfloat[Northern and Eastern Filament]{\label{a}\includegraphics[width=0.87\textwidth]{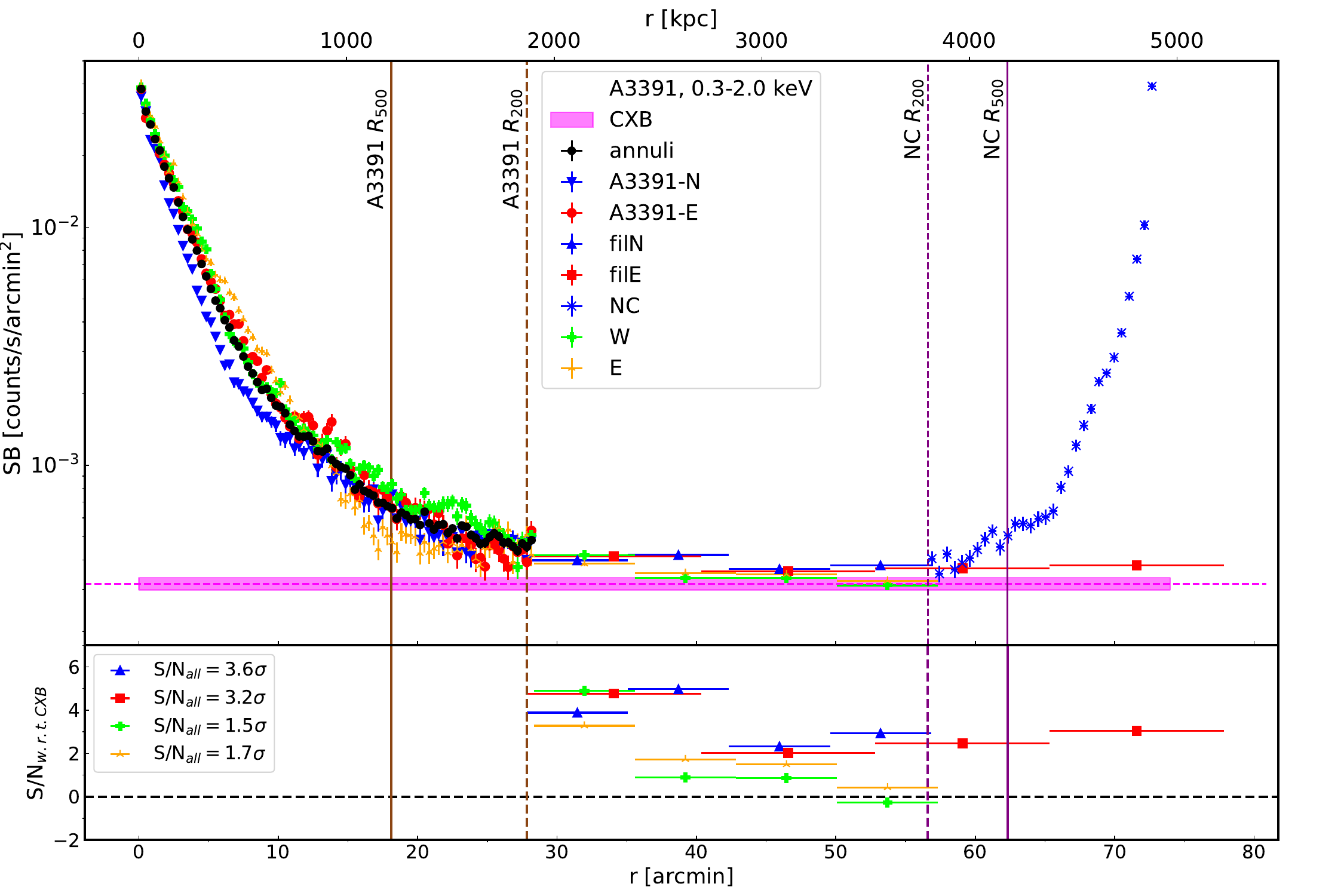}}\par
\subfloat[Southern Filament]{\label{b}\includegraphics[width=0.87\textwidth]{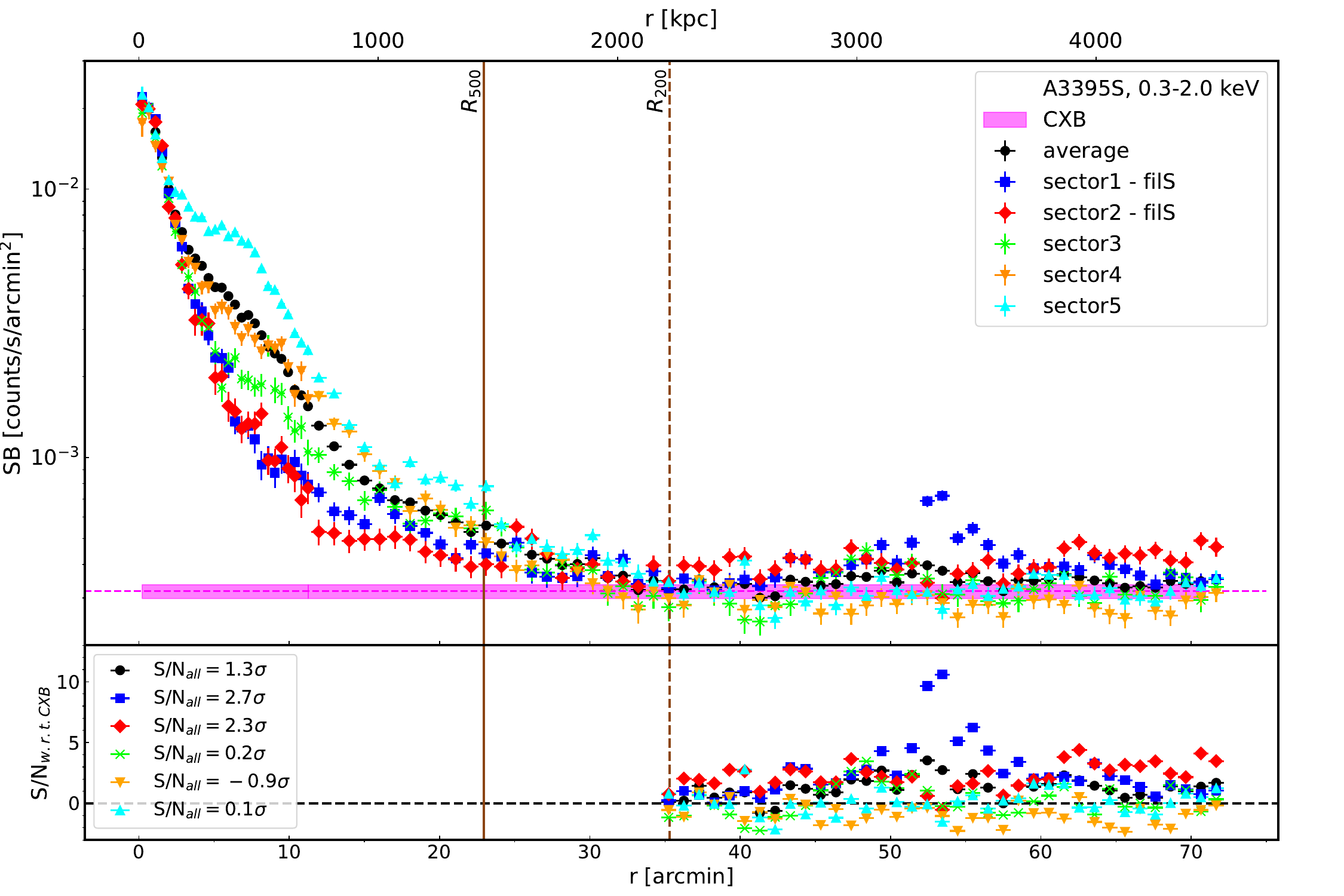}}
\caption{PIB-subtracted SB profiles in the $0.3-2.0~\mathrm{keV}$ band (top plots) and the significance values with respect to CXB for $r>R_{200}$ data points (bottom plots). The overall significance, $\mathrm{S/N}_{\mathrm{all}}$, of the excess emission in this regime is listed in the legends. \textit{a}: Profiles from A3391-N, NC, and the Northern Filament in blue; A3391-E and Eastern Filament in red; and west and east of A3391 in green and orange (see Fig.~\ref{fig:SBconfig}, left). \textit{b}: Profiles from sector1-5 of A3395S. Sector1 (blue) and 2 (red) are related to the Southern Filament (see Fig.~\ref{fig:SBconfig}, right). The CXB level and its $1\sigma$ range are plotted as the magenta horizontal dashed-lines and magenta shaded area. The $R_{500}$ ($R_{200}$) of each cluster is plotted as vertical solid (dashed) lines.}
\label{fig:SBprofiles}
\end{figure*}

We present the configuration of the source region used for the surface brightness analysis around A3391 and A3395S clusters in the left and right panels of Fig.~\ref{fig:SBconfig}, respectively. The corresponding PIB-subtracted surface brightness profiles in the $0.3-2.0~\mathrm{keV}$ band are shown in Fig.~\ref{fig:SBprofiles}a and b (top plots). In the bottom plots of Fig.~\ref{fig:SBprofiles}a and b, we show the significance of the surface brightness values for $r>R_{200}$ with respect to the CXB level ($\mathrm{S/N_{\mathrm{w.r.t. CXB}}}$). The CXB level and its standard deviation (see Sect.~\ref{sec:imag_analysis}) are shown as magenta dashed lines and the magenta shaded area. The overall significance of the surface brightness excess with respect to the CXB, $\mathrm{S/N}_{\mathrm{all}}$, of the different sectors in this regime is listed in the legend. The $\mathrm{S/N}_{\mathrm{all}}$ was calculated as follows:
\begin{equation}
    \mathrm{S/N}_{\mathrm{all}} = \frac{\mathrm{SB}_{r>R_{200}} - \mathrm{SB}_{\mathrm{CXB}}}{\sqrt{\sigma_{r>R_{200}}^2 + \sigma_{\mathrm{CXB}}^2}},
\label{eq:snr}
\end{equation}
where $\mathrm{SB}_{r>R_{200}}$ and $\sigma_{r>R_{200}}$ are the surface brightness value of the entire $r>R_{200}$ region and its statistical standard deviation, and $\mathrm{SB}_{\mathrm{CXB}}$ and $\sigma_{\mathrm{CXB}}$ are the sky background surface brightness and its standard deviation, which includes a systematic uncertainty (see Sect.~\ref{sec:imag_analysis}).
\par
In the directions of the Northern and Eastern Filament, we calculated A3391 sectorized surface brightness profiles out to $R_{200}$, namely A3391-N and A3391-E (cyan and red sectors in Fig.~\ref{fig:SBconfig}, left). The surface brightness values of the Northern (Eastern) Filament were then calculated from the four cyan (red) boxes of the same size, which were placed outside the $R_{200}$ of the A3391 cluster and the NC. Each Northern Filament box has a size of $7.25'\times 50'$ and is angled at $234.71^\circ$ from the $R_{200}$ of the A3391 in the NC direction. For the Eastern Filament, each box has the size of $12.5' \times 33.33'$ and is angled at $300^\circ$ from the $R_{200}$ of the A3391. In the Northern Filament direction, we continued with the surface brightness profile of NC, which was calculated from annuli around the center of the cluster. The data points related to the Northern Filament (filN) and Eastern Filament (filE) are shown respectively as blue upward triangles and red squares in Fig.~\ref{fig:SBprofiles}a. To compare with the surface brightness profiles from the filament regions, we also computed surface brightness profiles in the western (green) and eastern (orange) directions out to $57.33'$ ($\sim\!2R_{200}$) from the cluster center. The western and eastern sectors have position angles of $316^\circ$ to $378^\circ$ and $149^\circ$ to $209^\circ$, respectively. The two apparent extended sources at the edge of the eastern sector are background sources located at $z\sim0.1$, and therefore we excluded them from our analysis. Due to the influence of the A3395S/N cluster, we did not calculate any surface brightness profiles in the southern direction. The black data points in Fig.~\ref{fig:SBprofiles}a are the average surface brightness profile out to the $R_{200}$ of A3391 (excluding the southern area). By comparing the sectorized profiles with this average profile, we checked for any deviations (enhancement or depression) from the spherically symmetric surface brightness profile.
\par
As the A3395S/N cluster comprises two components, a different approach was taken to place the surface brightness regions compared to that used to place those around the A3391 cluster, such that we avoided treating the cluster as a single entity. To minimize the amount of emission from A3395N and to get the coverage of the Southern Filament, we placed five sectors covering the lower half of A3395S (Fig.~\ref{fig:SBconfig}, right). These sectors were centered at the A3395S (see Table~\ref{tab:clusters}) and have an opening angle of $36^\circ$ each, with the position angle from Sector1 to Sector5 spanning from $215^\circ$ to $395^\circ$. From each sector, we calculated the surface brightness profile out to $72.17'$ ($\sim\!2R_{200}$) from the center. The black data points in Fig.~\ref{fig:SBprofiles}b are the average surface brightness profile of all sectors.

\subsubsection{Inner region ($r\leq R_{500}$)}
As seen in Fig.~\ref{fig:SBprofiles}a (top plot), within $1-11'$, the surface brightness values of the A3391-N (blue downward-pointing triangles) are significantly lower than the other directions. In this radial range, compared with the full annuli profile (black data points), the A3391-N surface brightness values are lower, with relative differences ranging $8\%-36\%$. The lowest surface brightness value in this radial range is located at $6.3'$ with a significance of $11.5\sigma$ to the full annulus value. This shows that A3391 has an elliptical morphology, where the minor axis is in the north-south direction while the major axis is in the east-west direction.
\par
Similarly, for A3395S, we observe significantly higher surface brightness in Sector5 (Fig.~\ref{fig:SBprofiles}b, cyan upward-pointing triangles). Within $2.7-21.6'$, the relative difference values with respect to the average surface brightness profile (black data points) range between $16\%$ and $93\%$, with the peak located at $7.1'$ with a $13.7\sigma$ significance. This bump shows the apparent cluster elongation in this direction. From the center to $\sim\!15'$, the profiles of Sector1 (blue squares) and Sector2 (red diamonds) drop rapidly, and then flatten outward, denoting the pronounced shorter edge of the cluster in the southern direction.

\subsubsection{Outskirts region ($R_{500}<r<R_{200}$)}\label{sec3:surb_outs}
Within $R_{500}<r<R_{200}$, the enhancement in the western profile of A3391 (Fig.~\ref{fig:SBprofiles}a, green diamonds) is apparent. This is not unexpected given that the elongation of the cluster is in this direction. The average enhancement in this radial range with respect to the annuli profile (black data points) is $14\pm2\%$. For the other sectors, the surface brightness profiles are on average lower than the annuli with relative differences of $4\pm2\%$ (A3391-N), $3\pm2\%$ (A3391-E), and $12\pm2\%$ (E), meaning that the A3391-N and A3391-E are still brighter than the E sector. While ICM emission is still expected in this regime, other sources of emission, for example from the filaments, may contribute to this outskirts excess.
\par
For the A3395 profiles (Fig.~\ref{fig:SBprofiles}b) in the $R_{500}<r<R_{200}$ range, an enhancement is observed for Sector5 (cyan) with $15\pm2\%$ average excess with respect to the average profile in the same regime (black points). We attribute this enhancement to the emission from the cluster in this direction. Decrements with an average of $5\pm3\%$ and $9\pm2\%$ are observed for Sector3 (green) and Sector4 (orange). Similar to the case for A3391, insignificant decrements with respect to the average profile of $4\pm2\%$ and $2\pm3\%$ are seen for Sector1 (blue) and Sector2 (red).

\subsubsection{Filament region ($r>R_{200}$)}\label{sec3:surb_fils}
Beyond $R_{200}$, on average we observe higher surface brightness values in the filN and filE profiles for A3391 (blue upward-pointing triangles and red squares) than in the western (green) and eastern profiles (orange), indicating an enhancement in the intercluster regime. The overall significance values ($\mathrm{S/N}_{\mathrm{all}}$; Eq.~\ref{eq:snr}) for these $r>R_{200}$ regions show that the Northern and Eastern Filament are $>3.2\sigma$ away from the CXB level, while the Western and Eastern Sectors are $1.5-1.7\sigma$ away. We note that the first $r>R_{200}$ bins of all sectors are greater than $3.2\sigma$. In general, the enhancement near $R_{200}$ is expected to originate from the ICM. As can already be observed from the image (e.g., Fig.~\ref{fig:SBconfig}, left), there is an apparent elongation of the cluster to the west. Moreover, the surface brightness elevation of the eastern profile could result from additional residual emission from the excluded background bright galaxies, seen at $\sim\!28'$ and $50'$. Removing these first bins in the $r>R_{200}$ regime results in $\mathrm{S/N}_{\mathrm{all}}=3.5\sigma$ (filN), $2.5\sigma$ (filE), $0.5\sigma$ (W), and $1.2\sigma$ (E). This shows that there is still significant excess emission in the Northern and Eastern Filament regions in comparison to the other directions.
\par
For A3395S's $r>R_{200}$, a striking surface brightness peak is seen at around $54'$ ($\sim\!1.5R_{200}$) of Sector1. This peak is attributed to the LSC. For Sector1 and Sector2, the $\mathrm{S/N}_{\mathrm{all}}$ values in this regime are 2.7 and $2.3\sigma$, respectively. Combining Sector1 and Sector2, we obtain an excess with an overall significance of $2.5\sigma$. Meanwhile, for Sector3-5, the $\mathrm{S/N}_{\mathrm{all}}$ values are consistent with the CXB level, such that the $\mathrm{S/N}_{\mathrm{all}}$ values lie around zero (Fig.~\ref{fig:SBprofiles}b, bottom plot).
%-------------------------------------- 
%-------------------------------------- 
%-------------------------------------- 
\subsection{Spectral analysis}\label{sec3:spectro}
\subsubsection{The Little Southern Clump}\label{Sec:LSC}
\begin{figure}
\centering
\includegraphics[width=\columnwidth]{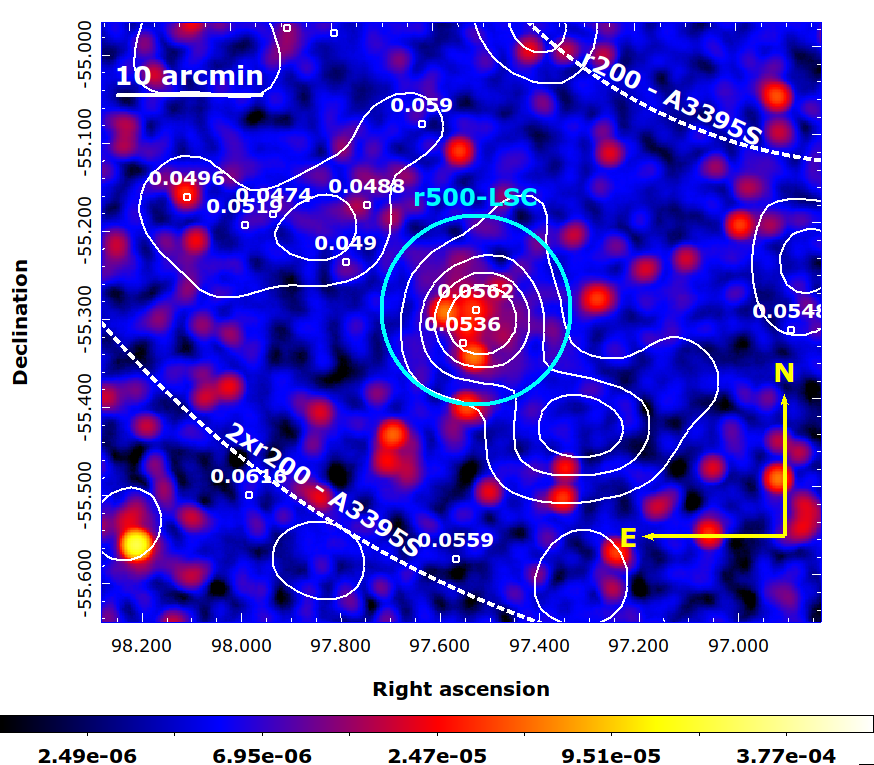}
\caption{eROSITA fully corrected count rate image in the $0.3-2.0$ keV band (smoothed with a $64''$ Gaussian kernel). The cyan circle indicates the $R_{500}$ of the LSC, centered at the coordinates of the WISEA J063004.80-551751.5 galaxy. The small white circles are known galaxies within $0.0424\le z \le 0.0636$ with their redshifts labeled. The overlaid white contours are from the optical DECam/DES galaxy density map. The white dashed lines mark the $R_{200}$ and $2R_{200}$ of the A3395S cluster.}
\label{fig:LSC}
\end{figure}

The LSC is an extended source that resides in the Southern Filament and has completely entered the $2R_{200}$ of the A3395 cluster, as shown in Fig.~\ref{fig:spectro_regions}. We show a zoomed-in image around the LSC in Fig.~\ref{fig:LSC}. We overlaid the image with the A3391/95 galaxy density map \citep[white contour; also see Sect. 2.2 of][]{Reiprich_2021} obtained from the Dark Energy Camera (DECam, \cite{DECam_2015}) as part of the Dark Energy Survey (DES, \cite{DES_2016}). The white circles mark the galaxies retrieved from the NED\footnote{The NASA/IPAC Extragalactic Database (NED) is funded by the National Aeronautics and Space Administration and operated by the California Institute of Technology.} with known redshifts and further selected with $0.0424\le z \le 0.0636$. We labeled their redshifts on top of each circle. The cyan circle indicates the calculated $R_{500}$ of the LSC and the white dashed lines mark the $R_{200}$ and $2R_{200}$ of the A3395S cluster.
\par
To characterize the LSC and to exclude it during the analysis of the Southern Filament, we need to first determine some characteristic properties of this object. Through spectroscopy, we acquired the $k_\mathrm{B}T_{500}$, which was then used to infer the $M_{500}$ and consequently, the $R_{500}$. The $k_\mathrm{B}T_{500}$ was calculated through an iterative procedure, such that we extracted and fit spectra from an annulus centered at the WISEA J063004.80-551751.5 galaxy at $(\alpha,\delta)=(6\! :\! 30\! :\! 04.8,\: -55\! :\! 17\! :\! 51.504)$. The inner and outer radii were varied until they correspond roughly to the $0.2-0.5R_{500}$ of the source region, where the $R_{500}$ values are listed in Table~\ref{tab:clusters}. Based on this, we obtain $k_\mathrm{B}T_{500}=0.94_{-0.07}^{+0.06}~\mathrm{keV}$ and $Z_{500}=0.10_{-0.04}^{+0.05}Z_\odot$ from an annulus with inner and outer radii of $1.3'$ and $3.2'$, respectively. We input this value into the mass$-$temperature ($M-T$) scaling relation by \cite{Lovisari_2015},

\begin{equation}
\log(M/C1) = a \cdot \log(T/C2) + b,
\label{eq:scaling}
\end{equation}
where $a = 1.65\pm0.07$, $b=0.19\pm0.02$, $C_1=5\times10^{13}h_{70}^{-1} M_{\odot}$, and $C_2=2.0$ keV. Assuming spherical symmetry and taking 500 times the critical density of the Universe at the WISEA J063004.80-551751.5 redshift as $\rho_{500}(z=0.0562)=4.51\times10^{-27}$ g cm$^{-3}$, we obtain $M_{500}=(2.33_{-0.34}^{+0.31})\times10^{13}M_{\odot}$ and $R_{500}=(6.44_{-0.32}^{+0.29})'\approx437.23_{-21.47}^{+19.38}~\mathrm{kpc}$. The calculated mass is lower than the one calculated using the eFEDS scaling relation \citep{Chiu_2022}, $(4.30\pm2.8)\times10^{13}M_{\odot}$, but is nevertheless in agreement within their $1\sigma$ uncertainties. As the studied group is closer to the cluster population considered in \cite{Lovisari_2015}, which are average- and low-mass systems at low-$z$, we use the $M_{500}=(2.33_{-0.34}^{+0.31})\times10^{13}M_{\odot}$ as the default.
\par
Based on the results, we classify the LSC as a small group of galaxies. This is further supported by the DECam/DES galaxy density map (Fig.~\ref{fig:LSC}, white contours), where galaxy overdensity was observed within the $R_{500}$ and extending towards the southwest. We also notice an overdensity in the northeast direction that coincides with six known galaxies. As has been discovered in simulations, galaxy groups that are within the vicinity of a cluster can accrete some field galaxies and grow through this process \citep{Vijayaraghavan_2013, Kuchner_2022}. However, these galaxies, together with the LSC, may also simply be on their way to being accreted towards the A3395 cluster with the filament. Deeper pointed observations, for example by \xmm and \chandra, will allow us to study the LSC and its surroundings in more detail. With the newly determined $R_{500}$ of the LSC, we performed two analyses of the Southern Filament, namely including and omitting the emission from the LSC within this radius (Sect.~\ref{sec3:spectro_fils}).

%-------------------------------------- 
\subsubsection{The outskirts ($R_{500}-R_{200}$)}\label{sec3:spectro_outs}
\begin{figure}
\centering
\includegraphics[width=\columnwidth,trim=1.7cm 1.1cm 1.7cm 2.1cm,clip]{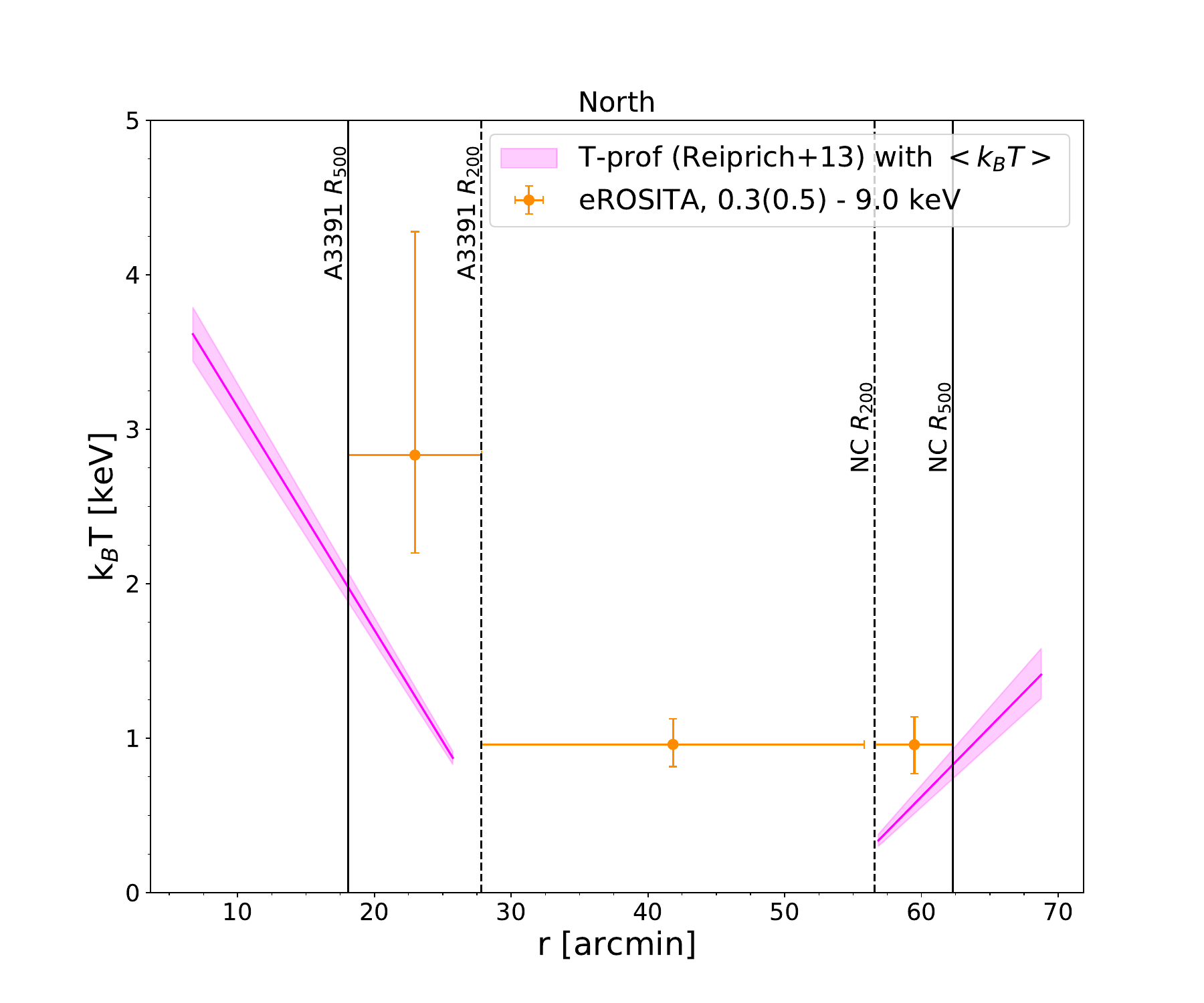}
\includegraphics[width=\columnwidth,trim=1.7cm 1.1cm 1.7cm 1.8cm,clip]{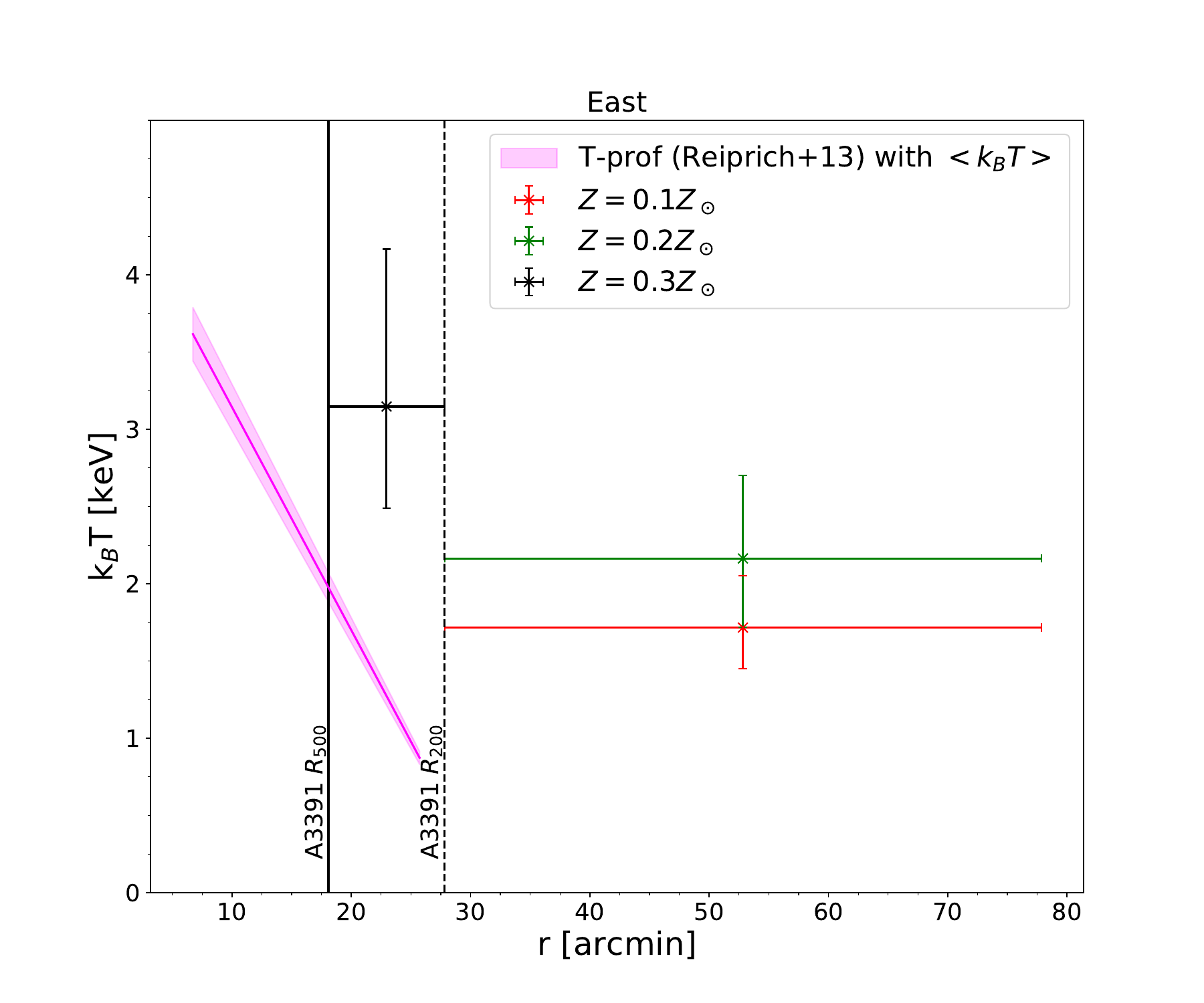}
\includegraphics[width=\columnwidth,trim=1.7cm 1.1cm 1.7cm 1.8cm,clip]{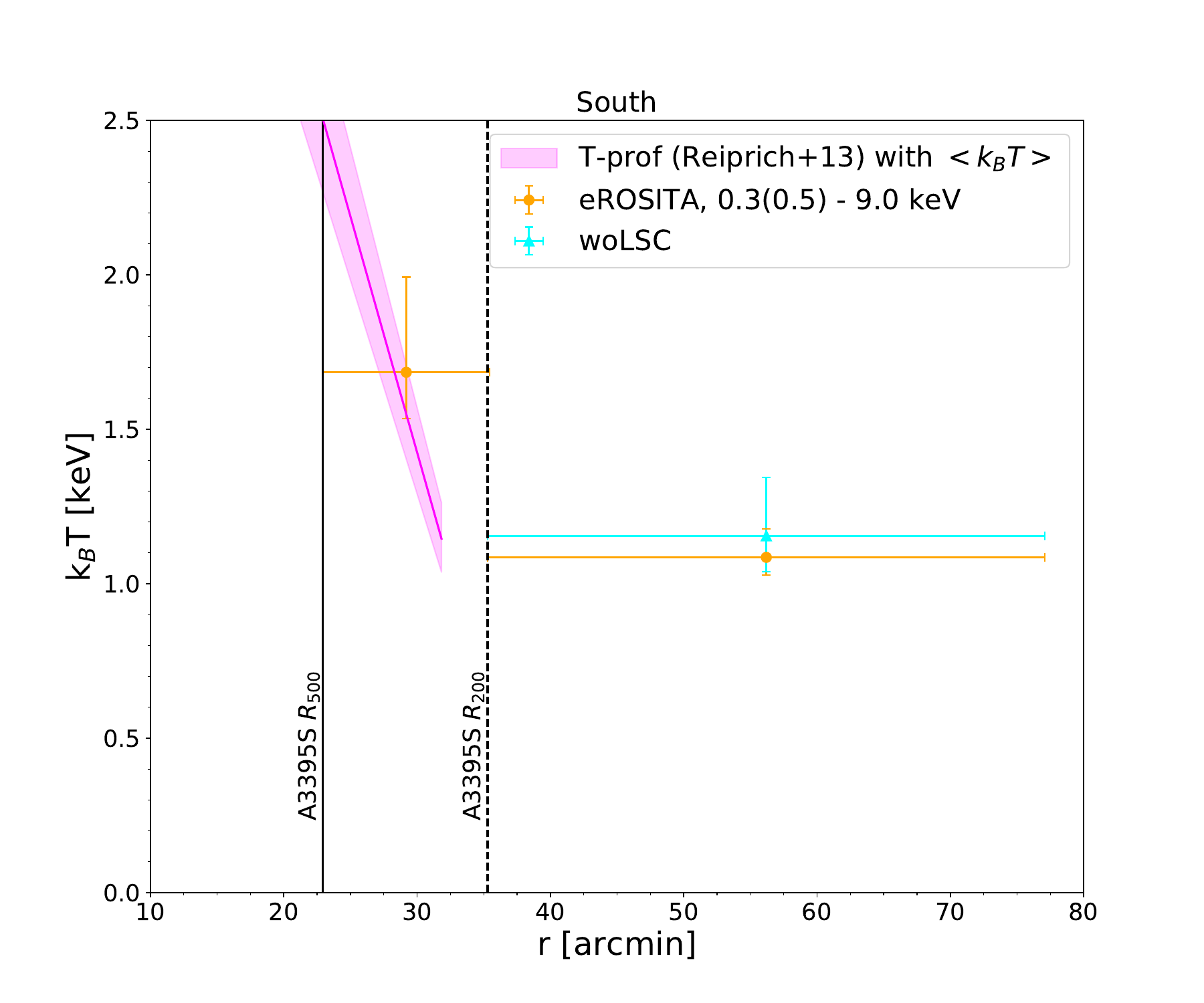}
\caption{eROSITA temperature profiles of the outskirts and filaments for the northern (top), eastern (middle), and southern (bottom) regions. The pink shaded areas are the $1\sigma$ temperature profiles from \cite{Reiprich_2013} using $k_\mathrm{B}T_{0.2-0.5R_{500}}$ as mean temperature of each cluster (see Table~\ref{tab:clusterTemps}). The black vertical solid (dotted) lines mark the $R_{500}$ ($R_{200}$) of the parent clusters.}
\label{fig:1Tprofiles}
\end{figure}

We list the results of the eROSITA spectral analysis in Table \ref{tab:spectro}. Starting from the northern outskirts of A3391 (A3391-N; Fig.~\ref{fig:spectro_regions}, white dashed lines sector), we report an upper limit of the metallicity of $0.31Z_\odot$. Meanwhile, due to a lack of statistics, we find that it is necessary to fix the metallicity to $0.3$ during the fit of the eastern sector (A3391-E; Fig.~\ref{fig:spectro_regions}, green sector). This is a typical metallicity value adopted for cluster outskirts (\citealp[e.g.,][]{Reiprich_2013, Urban_2017}). Overall, the normalizations and temperatures obtained from A3391-N and A3391-E are in good agreement with each other. The temperature values range between $2.2$ and $4.2~\mathrm{keV}$.
\par
The temperature of the outskirts of the NC (Fig.~\ref{fig:spectro_regions}, white dashed lines sector) is  $k_\mathrm{B}T=0.96_{-0.19}^{+0.18}~\mathrm{keV}$, which is lower than the temperature measured for the outskirts of the NC by \cite{Veronica_2022} based on data from \xmm, of namely $k_\mathrm{B}T=1.71_{-0.37}^{+0.66}~\mathrm{keV}$. Although there is overlap between the definitions of the source regions, they are not identical: the eROSITA region is placed in the south of the cluster at $10.62-16.34'$, while the \xmm source region is placed at $10.0-12.5'$. This temperature difference might also be the product of some other systematic differences, such as background treatment, point source selection, and/or instrumental calibration. Furthermore, the yielded metallicity is $Z=0.05_{-0.04}^{+0.06}Z_\odot$, whereas for \xmm the value has to be frozen at $0.3Z_\odot$ for statistical reasons. For the A3395 outskirts region (Fig.~\ref{fig:spectro_regions}, magenta box), the temperature and metallicity values are $1.68_{-0.15}^{+0.31}~\mathrm{keV}$ and $0.72_{-0.33}^{+0.60}Z_\odot$, respectively.
\par
Furthermore, we fit eROSITA spectra extracted over $0.2-0.5R_{500}$ of the A3391, A3395S/N, and NC clusters. The fitted eROSITA temperatures constrained using the $R_{500}$ ($k_\mathrm{B}T_{0.2-0.5R_{500}}$) from the literature and the eROSITA newly defined radii ($k_\mathrm{B}T_{0.2-0.5R_{500,eRO}}$), as well as other derived quantities ($M_{500,eRO}$ and $R_{500,eRO}$ are presented in Table \ref{tab:clusterTemps}. We report that the eROSITA temperatures for A3391 and A3395N are lower than \textit{ASCA} values listed in \cite{Reiprich_2002}, with significances of $2.4\sigma$ and $4.9\sigma$, respectively. The eROSITA best-fit temperature for A3395S is higher than that reported for A3391. We argue that since A3395S and A3395N are interacting, this hotter gas could be the result of some merging processes, for instance, shock heating or adiabatic compression. For NC, comparing with the \xmm temperature reported in \cite{Veronica_2022}, the eROSITA result is $25\%$ lower with $2.7\sigma$ significance. We note that this could either originate from some differences in the analyses, such as the source catalog used, or different background treatments or a calibration issue between the two instruments.
\par
To obtain the eROSITA temperatures ($k_\mathrm{B}T_{eRO}$) that were calculated within $0.2-0.5R_{500, eRO}$, we repeated the $M-T$ scaling relation procedure (see Sect.~\ref{Sec:LSC}), starting with the extraction of source spectra from an annulus region. We iterated the steps until the inner and outer radii of the source region corresponded to 0.2 and 0.5 of the calculated $R_{500, eRO}$. We acquire lower $k_\mathrm{B}T_{eRO}$ values, which consequently mean smaller eROSITA cluster radii  ($R_{500, eRO}$). If we consider these newly calculated eROSITA radii, for example $R_{500}=1.25R_{500, eRO}$ for A3391 cluster, and take into account our definition of inter-cluster emission, that is, that found at $r > R_{200}$, we find the filaments to be longer. However, for consistency, the characteristic radii used to present the results and discussion throughout this work (and other eROSITA A3391/95 papers) are those listed in Table~\ref{tab:clusters}, unless otherwise stated. These eROSITA cluster properties of the A3391/95 clusters are presented in Table~\ref{tab:clusterTemps}.
\par 
We employed the eROSITA $k_\mathrm{B}T$ values constrained from within apertures of $0.2-0.5R_{500}$ to compute the universal temperature profile from \cite{Reiprich_2013}. The profile describes the cluster temperatures within $0.3R_{200} < r < 1.15R_{200}$, such that $k_\mathrm{B}T=(1.19-0.84r/R_{200})\langle k_\mathrm{B}T \rangle$. We took $\langle k_\mathrm{B}T \rangle$ to be the eROSITA temperatures (Table~\ref{tab:clusterTemps}, second row). We display the temperature profiles of the outskirts and filament regions related to the Northern, Eastern, and Southern Filaments in the top, middle, and bottom panels of Fig.~\ref{fig:1Tprofiles}, respectively (data points). In each plot, the universal temperature profile from the corresponding cluster is plotted as the pink-shaded area. As shown, the acquired temperatures for the outskirts of A3391-N, A3391-E, and NC (Fig.~\ref{fig:1Tprofiles}, top and middle plots) are higher than the predicted values with a significance of $2.4\sigma$, $2.8\sigma$, and $1.6\sigma$, respectively.

\begin{table*}
    \centering
    \caption{eROSITA spectral analysis results, the derived electron density ($n_\mathrm{e}$), and the filament gas overdensity ($\delta_b$).}
    \resizebox{\textwidth}{!}{\begin{tabular}{c c c c c c c c}
    \hline
    \hline
\multirow{2}{*}{Regions} & \multirow{2}{*}{Fitting} & $norm$ & $k_\mathrm{B}T$ & $Z$ & $n_\mathrm{e}$ &  \multirow{2}{*}{$\delta_b$} & \multirow{2}{*}{stat/dof}\\
 & & $[10^{-6}$ cm$^{-5}/$arcmin$^2]$ & $[$keV$]$ & $[Z_{\odot}]$ & $[10^{-5}$ cm$^{-3}]$ & \\[5pt]
\hline \\[-1.7ex]
\hline \\[-1.7ex]
\multicolumn{8}{c}{\large{NORTH}}\\
\hline \\[-1.7ex]
\hline \\[-1.7ex]
A3391-N & & $2.47_{-0.24}^{+0.24}$ & $2.83_{-0.63}^{+1.45}$ & $<0.31$ &  &  & 9001.1/9613 \\[5pt]
\hline \\[-1.7ex]
NC & & $3.68_{-0.86}^{+1.10}$ & $0.96_{-0.19}^{+0.18}$ & $0.05_{-0.04}^{+0.06}$ & &  & 6660.5/6630 \\[5pt]
\hline \\[-1.7ex]
\multirow{1}{*}{Box1+2} & & $1.65_{-0.18}^{+0.19}$ & $0.96_{-0.14}^{+0.17}$ & $<0.01$ & $5.93_{-0.34}^{+0.34}$ & $225_{-13}^{+13}$ & 10855.0/10743\\[5pt]
\hline \\[-1.7ex]
\hline \\[-1.7ex]
\multicolumn{8}{c}{\large{EAST}}\\
\hline \\[-1.7ex]
\hline \\[-1.7ex]
A3391-E & fix $Z$ & $2.18_{-0.15}^{+0.15}$ & $3.15_{-0.66}^{+1.02}$ & $0.3$ & & & 8570.5/9127 \\[5pt]
\hline \\[-1.7ex]
\multirow{2}{*}{Box1+2}  & fix $Z$ & $0.82_{-0.09}^{+0.09}$ & $1.72_{-0.26}^{+0.33}$ & $0.1$ & $5.04_{-0.30}^{+0.28}$ & $191_{-11}^{+11}$ & 11318.2/11078\\[5pt]
  & fix $Z$ & $0.73_{-0.08}^{+0.08}$ & $2.16_{-0.45}^{+0.54}$ & $0.2$ & $4.76_{-0.29}^{+0.26}$ & $180_{-11}^{+10}$ & 11344.9/11078\\[5pt]
\hline \\[-1.7ex]
\hline \\[-1.7ex]
\multicolumn{8}{c}{\large{SOUTH}}\\
\hline \\[-1.7ex]
\hline \\[-1.7ex]
A3395 & &$0.59_{-0.17}^{+0.18}$ & $1.68_{-0.15}^{+0.31}$ & $0.72_{-0.33}^{+0.60}$ & & & 10277.6/10701\\[5pt]
\hline \\[-1.7ex]
\multirow{2}{*}{Box1+2} & & $0.74_{-0.12}^{+0.13}$ & $1.09_{-0.06}^{+0.09}$ & $0.10_{-0.04}^{+0.05}$ & $4.37_{-0.36}^{+0.36}$ & $166_{-14}^{+14}$  & 12756.4/11368\\[5pt]
 & woLSC & $0.39_{-0.12}^{+0.13}$ & $1.16_{-0.12}^{+0.19}$ & $0.11_{-0.07}^{+0.13}$ & $3.08_{-0.51}^{+0.46}$ & $117_{-19}^{+17}$  & 12620.0/11363\\[5pt]
\hline
\hline
    \end{tabular}}
    \label{tab:spectro}
\end{table*}
%JS: Personally, I would reduce the number of decimal places in this table and Table 5 due to the size of the error bars. 

\begin{table*}
    \centering
    \caption{eROSITA X-ray properties of the A3391/95 clusters/groups. The first row ($k_\mathrm{B}T_{lit}$) shows the literature temperature values. The second row ($k_\mathrm{B}T_{0.2-0.5R_{500}}$) lists the eROSITA temperatures over the $0.2-0.5R_{500}$, where the used $R_{500}$ are from Table~\ref{tab:clusters}. The third row ($k_\mathrm{B}T_{0.2-0.5R_{500,eRO}}$) lists the eROSITA temperatures over the $0.2-0.5R_{500,eRO}$, where $R_{500,eRO}$ is the eROSITA radius derived from the eROSITA mass $M_{500,eRO}$ obtained through the $M-T$ scaling relation.}
        \begin{tabular}{c c c c c c c}
        \hline
        \hline \\ [-1.7ex]
        \multicolumn{2}{c}{Parameters} & A3391 & A3395S & A3395N & NC & LSC \\
        \hline \\ [-1.7ex]
        \multicolumn{2}{c}{$k_\mathrm{B}T_{lit}^\dagger$} & $5.40\pm0.6^a$ & $5.0\pm0.3^a$ & $5.0\pm0.3^a$ & $1.99\pm0.04^b$ & $-$\\ [5pt]
        \hline \\ [-1.7ex]
        \multicolumn{2}{c}{$k_\mathrm{B}T_{0.2-0.5R_{500}}^\dagger$} & $3.86\pm0.19$ & $5.05_{-0.48}^{+0.52}$ & $3.19\pm0.21$ & $1.50_{-0.17}^{+0.18}$ & $-$ \\ [5pt]
        \hline \\ [-1.7ex]
        \multicolumn{2}{c}{$k_\mathrm{B}T_{0.2-0.5R_{500,eRO}}^\dagger$} & $3.79_{-0.18}^{+0.18}$ & $4.55_{-0.40}^{+0.47}$ & $3.76_{-0.30}^{+0.30}$ & $1.41_{-0.09}^{+0.23}$ & $0.94_{-0.07}^{+0.06}$ \\ [5pt]
        \hline \\ [-1.7ex]
        \multicolumn{2}{c}{$M_{500,eRO}^\star$} & $2.54_{-0.33}^{+0.33}$ & $3.42_{-0.62}^{+0.69}$ & $2.51_{-0.42}^{+0.42}$ & $0.45_{-0.06}^{+0.12}$ & $0.23_{-0.03}^{+0.03}$ \\ [5pt]
        \hline \\ [-1.7ex]
        \multirow{2}{*}{$R_{500,eRO}$} & $[']$ & $14.45_{-0.63}^{+0.62}$ & $17.06_{-1.03}^{+1.15}$ & $15.37_{-0.86}^{+0.86}$ & $8.79_{-0.36}^{+0.80}$ & $6.44_{-0.32}^{+0.29}$ \\ [5pt]
        & [kpc] & $978.84_{-42.93}^{+42.80}$ & $1071.72_{-64.66}^{+72.45}$ & $967.52_{-54.17}^{+53.86}$ & $546.14_{-22.19}^{+49.98}$ & $437.23_{-21.47}^{+19.38}$ \\ [5pt]
        \hline
        \hline
        \multicolumn{7}{l}{\footnotesize $^a$\textit{ASCA} temperatures \citep{Reiprich_2002}, $^b$\xmm temperature \citep{Veronica_2022}, $^\dagger$[keV], $^\star[10^{14}M_\odot]$}\\
        \hline
        \hline
        \end{tabular}
\label{tab:clusterTemps}
\end{table*}

%-------------------------------------- 
\subsubsection{The filaments}\label{sec3:spectro_fils}
We used the obtained \texttt{apec} normalizations to derive the electron densities, $n_\mathrm{e}$, of the filament regions. We assumed a simple geometry for the filament, that is a cylinder with its axis in the plane of the sky. The $n_\mathrm{e}$ was calculated by rewriting Eq. \ref{eq:norm}:

\begin{equation}
\begin{split}
    n_\mathrm{e} = &\left[1.52\times10^{-10}~\mathrm{cm}^{-1}\times norm\times(1+z)^2 \right.\\
& \left. \times \left(\frac{D_A}{\mathrm{Mpc}}\right)^2 \times \left(\frac{r}{\mathrm{Mpc}}\right)^{-2} \times \left(\frac{h}{\mathrm{Mpc}}\right)^{-1} \right]^\frac{1}{2},
\end{split}
\label{eq:n_e}
\end{equation}
where, for instance, $D_A(z=0.0533) = 222.22~\mathrm{Mpc}$. The radius and height of the cylinder are represented by $r$ and $h$. The values of $r$ and $h$ of the filament boxes were introduced in Sect.~\ref{sec:spectro}. The hydrogen density is taken to be $n_\mathrm{H} \approx n_\mathrm{e}/1.17$. Given that the radii of our filament cylinders ranges from 1.07 to 1.62 Mpc, and that the gas in filaments is isothermal up to $\sim\!1.5~\mathrm{Mpc}$ from their cores, as reported by \cite{Galarraga_2021}, we deem our assumptions for calculating $n_\mathrm{e}$ to be reasonable.
\par
Using $n_\mathrm{e}$, we derived the filament gas density, $\rho_\mathrm{gas}$, that is
\begin{equation}
    \rho_\mathrm{gas} = \mu_\mathrm{e} \times n_\mathrm{e} \times m_\mathrm{p},
\label{eq:gas_dens}
\end{equation}
where $\mu_\mathrm{e} \approx 1.14$ is the mean molecular weight per electron, and $m_\mathrm{p} = 1.6726\times10^{-24}~\mathrm{g}$ is the proton mass. Then, we calculated the gas overdensity $\delta_b = \rho_\mathrm{gas}/\overline{\rho_b}(z)$, where $\overline{\rho_b}(z)=\Omega_b(z)\rho_\mathrm{cr}(z)$ is the mean baryon density at redshift $z$. With the assumed cosmology used in this work \citep{Planck2018_2020}, we obtained a baryon density at $z=0.0533(0.0525)$ of $\Omega_b=0.0561(0.0560)$, and a corresponding critical density of the Universe of $\rho_\mathrm{cr}=8.9861(8.9790)\times10^{-30}~\mathrm{g~cm^{-3}}$. We therefore have $\overline{\rho_b}(0.0533)=5.0455\times10^{-31}~\mathrm{g~cm^{-3}}$ and $\overline{\rho_b}(0.0525)=5.0320\times10^{-31}~\mathrm{g~cm^{-3}}$.
\par
To account for the systematic errors introduced by the assumed geometrical shape, we also calculated the $n_\mathrm{e}$ and $\delta_b$ using two alternative shape assumptions. The first shape is an elliptical cylinder, where the major axis and the height are assumed to be the same as the radius and the height of the previously assumed cylinder, while the minor axis is taken to be half of the radius. The second assumption is a cuboid shape, where we take the width of the base and the height to be the diameter and the height of the cylinder. The density contained in the elliptical cylinder shape is $41\%$ higher than that of the cylinder, while the density of the cuboid shape is $11\%$ lower. Using either of the geometrical shape assumptions does not change the classification of the A3391/95 filaments based on the overdensity criteria stated in \cite{Shull_2012}, namely that they fall between the WHIM criteria. In Table \ref{tab:spectro}, we therefore only list the $n_\mathrm{e}$ and $\delta_b$ values calculated whilst assuming the common cylindrical shape.
\par
We compared our results to the WHIM properties predicted by cosmological hydrodynamic simulations (\citealp[e.g.,][]{Dave_2001, Martizzi_2019}. \citealp[For observational constraints of WHIM properties, see, e.g.,][]{Shull_2012, Nicastro_2018, Kovac_2019}). Here, we adopted temperature cut of $T=10^5-10^7~\mathrm{K}~(k_\mathrm{B}T=0.009-0.862~\mathrm{keV})$ and the gas overdensity $\delta_b = 0.001-316.22$ for the WHIM \citep{Shull_2012}. Whereas, gas with $k_\mathrm{B}T>0.862~\mathrm{keV}$ and any number density $n_\mathrm{H}$ was considered as the hot medium (HM), which is the shock-heated gas in and near the most massive dark matter haloes \citep{Martizzi_2019}.
\par
We present the results of the spectral analysis of the different filaments in Table~\ref{tab:spectro}. Firstly, for the Northern Filament, we obtain a best-fit temperature of $\sim\!1~\mathrm{keV}$ and an upper metallicity limit of $0.01Z_\odot$. To investigate whether this low metallicity is caused by a bias of some kind, for example, by multi-temperature components, we performed a two-temperature (2T) component fit. We obtain a cool and a warm component of 0.1 and 1.3 keV, respectively. The cool component temperature is well within the range of the expected WHIM temperatures. The temperature of the warm component, on the other hand, agrees well with the 1T fit temperature and gives a higher metallicity upper limit of $0.12Z_\odot$. However, through the Bayesian information criterion \citep[BIC,][]{BIC} and the Akaike information criterion \citep[AIC,][]{AIC} tests, we find that adding a second thermal emission component to the model does not improve the fit. We note that deeper data are required to differentiate between the two models. We list the result of the two-temperature fitting in Table~\ref{tab:spectro_tests} of Appendix~\ref{app:spectrotests}. The derived gas overdensity, $\delta_b$, of the Northern Filament is $225\pm13$, which is within the predicted WHIM overdensities.
\par
For the Eastern Filament, we needed to fix the metallicity during the fit to typical values found for filaments from observations and simulations (\citealp[e.g.,][]{Tanimura_2020, Biffi_2022}), that is 0.1 and $0.2Z_\odot$. The average temperature of both fits is $1.9~\mathrm{keV}$ and the best-fit values of these fits are always within $1\sigma$ of one another. We note that fixing the metallicity has an impact on the normalization, such that we observe an anti-correlation between the value fixed for the metallicity and the resulting normalization. The $\delta_b$ from both metallicity fits ranges between 169 and 202.
\par
From the Southern Filament boxes, we obtain a $1.1~\mathrm{keV}$ gas with a metallicity value of $\sim\!0.10Z_\odot$. The eROSITA data enable us to identify the LSC as a clump within the filament, therefore allowing us to assess the influence of the clump on the Southern Filament (Sect.~\ref{Sec:LSC}). We performed spectral fittings excluding the LSC emission within its $R_{500}$ (hereafter, woLSC). The obtained best-fit temperature and metallicity are $1.2~\mathrm{keV}$ and $\sim\!0.11Z_\odot$, which means that we observe no significant changes in these parameters when including or excluding the LSC. However, the normalization per unit area drops by 47\%, and the gas overdensity decreases by 30\% (from $\delta_b=166\pm14$ to $117_{-19}^{+17}$) with a significance of $2.2\sigma$. Through this analysis, we directly demonstrate the extent to which the filament density is overestimated when ignoring the presence of a dense clump (\citealp[e.g.,][]{Nagai_2011, Eckert_2015clump, Mirakhor_2021}).

%-------------------------------------- 
\section{Discussion}\label{sec:discuss}
\subsection{Inner ($r\leq R_{500}$)}
Fig.~\ref{fig:SBprofiles} shows the surface brightness profiles of the A3391 (top) and A3395S (bottom) clusters from their center out to $\sim\!2R_{200}$. From the inner region of the profiles ($r<R_{500}$), we observe that the clusters are not spherically symmetric, but are rather elliptical. The elongation of the A3391 cluster is apparent in the east$-$west direction, while the A3395S cluster is extended in the northwest direction. A recent study employing galaxy groups and clusters at $z=0$ from large hydrodynamical simulations shows a correlation between their mass and morphology, as well as a correlation between their morphology and the number of filaments connected (connectivity) to them \citep{Gouin_2021}. In these simulations, the unrelaxed groups and clusters of high connectivity, such as the A3391 cluster, are more elliptical than their relaxed and weakly connected counterparts.

\subsection{Outskirts ($R_{500}<r<R_{200}$)}
For the outskirt regions ($R_{500}<r<R_{200}$), we observe higher surface brightness values in the elongation directions of the clusters (Fig.~\ref{fig:SBconfig} and \ref{fig:SBprofiles}). Despite this, the filament-facing sectors are still brighter than their non-filament-facing counterparts (see Sect.~\ref{sec3:surb_outs}). For instance, A3391-N and A3391-E of the A3391 cluster are brighter than the E Sector, and Sectors 1 and 2 of the A3395S cluster are brighter than Sectors 3 and 4.
\par
The spectral analysis (Sect.~\ref{sec3:spectro_outs} and Table~\ref{tab:spectro}) suggests that the outskirts temperatures of A3391-N, A3391-E, and NC are higher than the universal temperature profile (\cite{Reiprich_2013}; Fig.~\ref{fig:1Tprofiles}, pink shaded areas) with a significance of $2.4\sigma$, $2.8\sigma$, and $1.6\sigma$, respectively. A similar comparison to the outskirts of the A3391 and NC clusters reveals a temperature enhancement in the northeastern outskirts of the A1689 cluster \citep{Kawaharada_2010}. These latter authors speculated that this is due to the thermalization process induced by the filament in the mentioned direction.
\par
The constrained metallicity values in the outskirts of A3391 and NC (excluding A3391-E, where the metallicity is frozen) range between 0.1 and $0.3Z_\odot$. This is in good agreement with other measurements; for example, \textit{Suzaku} measurements of the outskirts of the Perseus cluster \citep{Simionescu_2011}, the A3112 cluster \citep{Ezer_2017}, and ten other nearby galaxy clusters \citep{Urban_2017}, as well as the cosmological hydrodynamical simulations (\citealp[e.g.,][]{Biffi_2018}). The metallicity measurements from the observational data reported in the literature are constrained from various radial bins out to $R_{200}$. In the regime defined as outskirts in our work ($R_{500}<r<R_{200}$), the reported metallicities in the literature are consistent with $\sim0.3Z_\odot$, with large uncertainties ranging between 0.1 and $0.6Z_\odot$. The metallicity of the outskirts of A3395 is high ($0.72_{-0.33}^{+0.60}Z_\odot$), but is still broadly in agreement with the literature values. The rather high metal content found in the outskirts is supposedly the result of the accretion of already enriched gas together with more pristine gas during the cluster assembly \citep{Ezer_2017, Biffi_2018}. Metal production occurs at higher redshifts ($z \approx 2-3$) through some supernova (SN) events (SN Type II and SN Type Ia) and asymptotic giant branch (AGB) stars (\citealp[e.g.,][]{Werner_2008metal, Nomoto_2013}) and this material is later mixed and distributed throughout the ICM by various processes, such as AGN feedback, galactic winds, or ram-pressure stripping (\citealp[e.g.,][]{Schindler_2010, Ettori_2013}).

\subsection{Filaments ($r>R_{200}$)}
From the surface brightness analysis (Sect.~\ref{sec3:surb_fils}), we calculated the significance of the emission of all directions in the $r>R_{200}$ regime with respect to the CXB level ($\mathrm{S/N}_{\mathrm{all}}$; Eq.~\ref{eq:snr}). We confirmed the surface brightness excess of the detected filaments \citep{Reiprich_2021}. For the A3391 cluster, we obtain $\mathrm{S/N}_{\mathrm{all}}=3.5\sigma$ and $2.5\sigma$ for the Northern and Eastern Filament, respectively, and $0.5\sigma$ and $1.2\sigma$ for the western and eastern sectors (non-filament-facing directions). For A3395S, excesses with a significance of $\mathrm{S/N}_{\mathrm{all}}=2.7\sigma$ and $2.3\sigma$ are acquired from the filament sectors, while the surface brightness of the other sectors is around the CXB level ($\mathrm{S/N}_{\mathrm{all}}\approx0\sigma$). While the significance of the excess emission of the Northern and Eastern Filaments is in good agreement with that reported in \citet[][see their Table 5]{Reiprich_2021}, the significance of the excess emission of the Southern Filament is $\sim\!1.5\sigma$ lower. As our goal here is to construct surface brightness profiles, we do not have identical surface brightness regions for the filaments as defined in \cite{Reiprich_2021} and therefore small differences are expected.
\par
We characterized the properties of the filament gas through spectral analysis (Sect.~\ref{sec3:spectro_fils}). While the gas overdensity of our filaments ($98\leq\delta_b\leq 238$) falls within the predicted WHIM gas overdensity, the temperatures obtained for both the Northern and Southern Filament ($k_\mathrm{B}T\approx1.0~\mathrm{keV}$) are more consistent with the HM phase. Nevertheless, these temperatures are close to the upper limit of the WHIM temperature ($\sim\!0.9~\mathrm{keV}$, as adopted in the simulations of \citealt{Shull_2012}). Our findings for temperature and density are in accordance with the hot gas phase of filaments in the IllustrisTNG simulation (\citealp[lower right panel of Fig. 6 and 8 of][]{Galarraga_2021}). These latter authors found that the profiles of the hot gas phase of the filaments are flat with an average temperature of below $2\times10^7$ K ($\sim1.7$ keV), while the density profiles start at $\sim\!10^{-4}~\mathrm{cm}^{-3}$ and drop to $\sim\!10^{-5}~\mathrm{cm}^{-3}$ at $\sim\!1$ Mpc from the filament spines. We recall that the radial widths of the A3391/95 filaments in this work are $r=0.45$ Mpc for Northern Filament and 1.33 Mpc for Southern Filament, and their lengths are $1.8~\mathrm{Mpc}$ and $2.7~\mathrm{Mpc}$, respectively, which means they are considered to be short filaments\footnote{Based on the filament populations defined in \cite{Galarraga_2021}, such as short ($L_f<9~\mathrm{Mpc}$), medium ($9\leq L_f<20~\mathrm{Mpc}$), and long ($L_f\geq20~\mathrm{Mpc}$).}. The shorter filaments usually trace a denser large-scale environment and are therefore more likely to be associated with the over-dense structures rather than the skeleton of the cosmic web \citep{Galarraga_2020, Galarraga_2021, Galarraga_2022, Vurm_2023}. Furthermore, they are in deeper potential wells and experience stronger gravitational heating. Our temperatures and gas overdensities are also within the ranges found for gas at $r>R_{200}$ in the EAGLE hydrodynamical simulation \citep{Tuominen_2021}. From comparison with other observations, we see that the temperatures of our filaments are consistent with the filament temperatures reported in \cite{Tanimura_2020} from the ROSAT stacked analysis of $15\,165$ filaments ($0.9_{-0.6}^{+1.0}$ keV) and with those reported in \cite{Tanimura_2022} from the eFEDS stacked analysis of 463 filaments ($1.0_{-0.2}^{+0.3}$ keV). \cite{Tanimura_2020, Tanimura_2022} restricted their analyses to a radial range of $r<2$ Mpc, and the radial widths of the A3391/95 filaments in this work fall within this range.
\par
We find very good agreement between our findings and the gas properties of the simulated A3391/95 analog clusters (outside of their $R_{200}$) of \cite{Biffi_2022}. As depicted in their gas phase-space diagram (Fig. 9b), these authors find the bridge gas to be in the warm-hot phase with a typical temperature of $\sim\!1~\mathrm{keV}$ and a median overdensity of $\sim\!100$ \citep{Biffi_2022}.
\par
Furthermore, we find that the Northern and Southern Filaments are poorly enriched by metals; this is expected in these low-density regions as they are further away from any metal production sites, such as star-forming regions \citep{Biffi_2022}. While the metallicity values for these filaments are unexpectedly low, these values are not completely excluded by the simulations (\citealp[e.g., Fig. 9b of][]{Biffi_2022}). Deeper data is required to confirm the low metallicity value.
\par
The gas temperature of the Eastern Filament ranges between 1.5 and 2.7 keV. Nevertheless, as reported by \cite{Reiprich_2021}, the significance of the detection of the Eastern Filament is less than $3\sigma$, which is tentative. We note that the Northern and Southern Filament are confirmed in the Planck-SZ and DECam galaxy density maps, as shown in \cite{Reiprich_2021}. From the spectral modeling, we obtain $3.6-4.4\sigma$ significance for the Northern and Southern Filament and $<3\sigma$ significance for the Eastern and Southern Filament without LSC. As the source regions used and the number of the free parameters in the imaging and spectral analyses differ, small differences in the values obtained from both analyses are not unexpected.
\par
We performed additional tests on the outskirts regions of A3391 and A3395, the Northern Filament, and Southern Filaments. These tests involved treating the filament boxes as a possible sky foreground. The results and discussion of these tests can be found in Appendix~\ref{app:spectrotests}.
%-----------------------------------------------------------------
%-----------------------------------------------------------------
\section{Summary and conclusions}\label{sec:summarize}
We investigated the outskirts ($R_{500} < r < R_{200}$) and the detected inter-cluster filaments ($r> R_{200}$) of the A3391/95 system using the eROSITA PV data. We focused our analysis on the northern and northeastern directions of A3391, and the southern direction of A3395, which are in the directions of the Northern, Eastern, and Southern Filaments. We created images to generate a PIB-subtracted, exposure-, and Galactic-absorption-corrected image in the soft energy band of $0.3-2.0~\mathrm{keV}$. Using the final corrected products, we calculated surface brightness profiles out to $\sim\!2R_{200}$ of the A3391 and A3395 clusters in various directions. We performed comprehensive spectral analyses in the outskirts and filament regions. We constrained the gas properties, including the normalizations, temperatures, and metallicities. Utilizing the acquired normalizations and assuming a cylinder shape with its axis in the plane of the sky, we derived other quantities, namely the electron densities ($n_\mathrm{e}$) and gas overdensities ($\delta_b$). We compared our results with simulated filament properties. We summarize our findings below:
\begin{itemize}
    \item The X-ray surface brightness profiles at $r<R_{500}$ in various directions emphasize the morphology of A3391 and A3395S. We notice significant surface brightness decrements below $10'$ in the northern direction of A3391 and higher values below $22'$ in the northwestern profile of A3395S. These show the apparent ellipticity of A3391 in the east$-$west direction and the extension of A3395S in the northwestern direction, respectively.
    \item The temperatures in the filament-facing outskirts of A3391-N, A3391-E, and NC are higher than the temperatures expected for a typical cluster outskirt profile predicted using the universal temperature profile of \cite{Reiprich_2013}, with significances of $2.4\sigma$, $2.8\sigma$, and $1.6\sigma$, respectively. These enhancements may be related to heating processes induced by the filaments.
    \item We confirm surface brightness excesses in the profiles of the Northern, Eastern, and Southern Filaments. In the surface brightness profile of the Southern Filament, we observe a peak at $\sim\!1.5R_{200}$ due to the LSC. 
    \item We detect hot gas beyond the $R_{200}$ of the cluster (filament regions).
    The Northern Filament has a best-fit temperature of $0.96_{-0.14}^{+0.17}~\mathrm{keV}$, while the Southern Filament has a best-fit temperature of $1.09_{-0.13}^{+0.06}~\mathrm{keV}$. The gas overdensities of the Northern and Southern Filaments are found to be within the ranges of $212< \delta_b < 237$ and $152 < \delta_b < 180$, respectively. The density of these filaments is within the expected range of the simulated WHIM properties. The enhanced temperatures of both filaments may be due to the fact that they are shorter and are located in a denser environment; consequently, they may experience stronger gravitational heating \citep{Galarraga_2020}. However, these values are close to the upper limit of WHIM temperatures ($\sim\!0.9~\mathrm{keV}$, as adopted in the simulations of \citealt{Shull_2012}). More detailed comparisons to the simulated filaments of similar lengths and environments are required, as well as comparisons to other systems in observations.
    \item An upper metallicity value of $<0.01Z_\odot$ is obtained for the Northern Filament and $0.10_{-0.04}^{+0.05}Z_\odot$ for the Southern Filament.
    \item The eROSITA data allowed us to identify the LSC, a clump within a filament. We characterized the properties of the LSC, such that $k_\mathrm{B}T_{500}=0.94_{-0.07}^{+0.06}~\mathrm{keV}$, $Z_{500}=0.10_{-0.04}^{+0.05}Z_\odot$, $M_{500}=\left(2.33_{-0.34}^{+0.31}\right)\times10^{13}M_{\odot}$, and $R_{500}=\left(6.44_{-0.32}^{+0.29}\right)'\approx437.23_{-21.47}^{+19.38}~\mathrm{kpc}$. From the optical DECam/DES galaxy density map, we observe some galaxy overdensities around the LSC. These galaxies may be being accreted towards the LSC, or they are simply a part of the filament and are on the way to the A3395 cluster. Deeper observations will enable us to study the LSC and its surroundings.
    \item In this work, we directly show the extent to which the filament gas density is overestimated when ignoring the presence of a clump. For example, when excluding the LSC from the analysis of the Southern Filament, we observe a decrease of 47\% in the normalization per unit area and 30\% in the gas overdensity with a significance of $2.2\sigma$. No significant changes to the temperature and metallicity of the gas were observed.
\end{itemize}

The outskirts of galaxy clusters are of paramount importance not only for locating large-scale structures but also for studying various accretion physics phenomena, such as shocks, mergers, and clumping. However, due to the faint nature and their location, studying the outskirts is most often hindered by a lack of sensitivity in the soft energy band and/or the limited FoV of the instruments. eROSITA, equipped with wide FoV, scan, and survey observation modes, and superior soft energy response, is an outstanding instrument with which to probe these regions. Indeed, through the A3391/95 PV observations, eROSITA has already demonstrated its capability to capturing these nearby clusters beyond their virial radii. Moreover, these observations reveal the filamentary structure that connects at least five galaxy groups and clusters. Here, we were able to go one step further and characterize the filament properties in terms of density, temperature, and metallicity. With the eROSITA all-sky survey (eRASS), further such systems will be found and this will improve the statistics of observational studies of inter-cluster filaments.

\begin{acknowledgements}
      The authors would like to thank the anonymous referee for their constructive feedback and suggestions that helped improve the presentation of the manuscript.
      We thank Gabriele Ponti for the valuable discussion about the eROSITA foreground.
      Funded by the Deutsche Forschungsgemeinschaft (DFG, German Research Foundation) – 450861021.
      This research was supported by the Excellence Cluster ORIGINS which is funded by the Deutsche Forschungsgemeinschaft (DFG, German Research Foundation) under Germany's Excellence Strategy – EXC-2094 – 390783311.
      VB acknowledges funding by the Deutsche Forschungsgemeinschaft (DFG, German Research Foundation) --- 415510302.
      AV is a member of the  Max-Planck International School for Astronomy and Astrophysics (IMPRS) and of the Bonn-Cologne Graduate School for Physics and Astronomy (BCGS), and thanks for their support.
      KD acknowledges support by the COMPLEX project from the European Research Council (ERC) under the European Union’s Horizon 2020 research and innovation program grant agreement ERC-2019-AdG 882679 and by the Deutsche Forschungsgemeinschaft (DFG, German Research Foundation) under Germany’s Excellence Strategy - EXC-2094 - 390783311.
      CS and TR acknowledge support from the German Federal Ministry of Economics and Technology (BMWi) provided through the German Space Agency (DLR) under project 50 OR 2112.
      This work is based on data from eROSITA, the soft X-ray instrument aboard SRG, a joint Russian-German science mission supported by the Russian Space Agency (Roskosmos), in the interests of the Russian Academy of Sciences represented by its Space Research Institute (IKI), and the Deutsches Zentrum für Luft- und Raumfahrt (DLR). The SRG spacecraft was built by Lavochkin Association (NPOL) and its subcontractors, and is operated by NPOL with support from the Max Planck Institute for Extraterrestrial Physics (MPE). The development and construction of the eROSITA X-ray instrument was led by MPE, with contributions from the Dr. Karl Remeis Observatory Bamberg and ECAP (FAU Erlangen-Nuernberg), the University of Hamburg Observatory, the Leibniz Institute for Astrophysics Potsdam (AIP), and the Institute for Astronomy and Astrophysics of the University of Tübingen, with the support of DLR and the Max Planck Society. The Argelander Institute for Astronomy of the University of Bonn and the Ludwig Maximilians Universität Munich also participated in the science preparation for eROSITA. The eROSITA data shown here were processed using the eSASS software system developed by the German eROSITA consortium.
      This research has made use of the NASA/IPAC Extragalactic Database (NED), which is funded by the National Aeronautics and Space Administration and operated by the California Institute of Technology.
\end{acknowledgements}

\bibliographystyle{aa}
\bibliography{list_bib}
%-------------------------------------------------------------------
\begin{appendix}
%%%%%%%%%%%%%%%%%%%%%%%%%%
\onecolumn

\section{Total $N_\mathrm{H}$ map of the A3391/95 field}\label{App:A}
\begin{figure}[h!]
\centering
\includegraphics[width=\columnwidth]{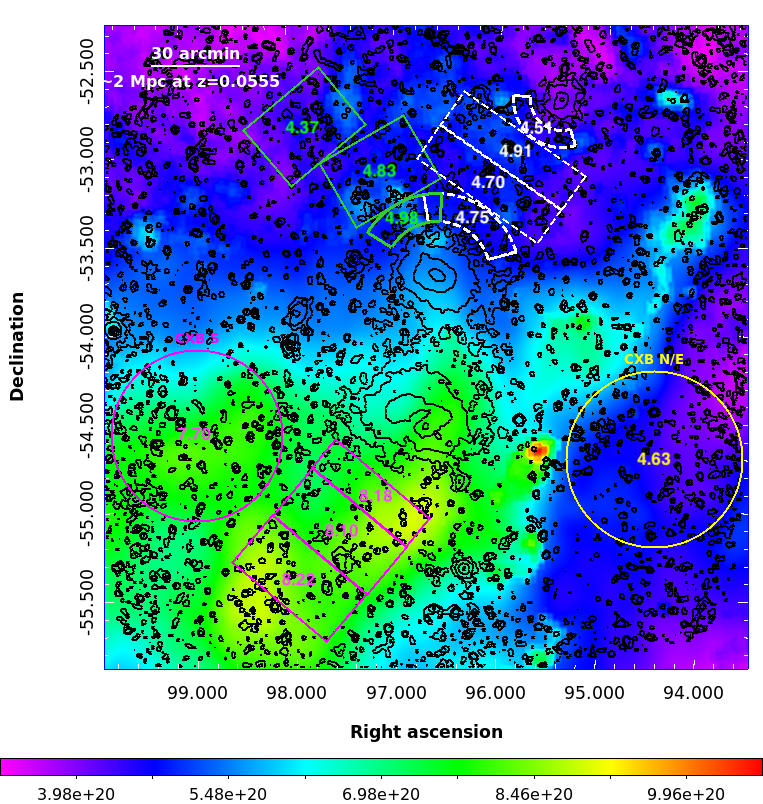}
\caption{The A3391/95 total $N_\mathrm{H}$ map generated from IRAS $100~\mathrm{\mu m}$ and HI4PI. We overlaid the source and CXB regions used for the spectral analysis (boxes, sectors, and circles). The numbers in each region are the average values in the units of $10^{20}~\mathrm{atoms~cm^{-2}}$. The eROSITA contours are plotted in black. The colorbar is in the units of $\mathrm{atoms~cm^{-2}}$.}
\label{fig:NH}
\end{figure}

\newpage
\section{Additional spectral analysis tests}\label{app:spectrotests}
The first test was to treat the Northern Filament (NF) and Southern Filament without LSC (SFwoLSC) as possible sky background regions. We used these filament regions to estimate the sky background components for the outskirts of the A3391-N and A3395 (A3391-N BGNF and A3395 BGSFwoLSC, respectively), thereby effectively assuming the filament emission as a foreground structure. We find that the normalization of both outskirts regions is reduced by a factor of $\sim\!2-3$. For the A3391-N BGNF, we can only constrain a lower limit temperature of 4.5 keV, while for the A3395 BGSFwoLSC, the temperature is of the same order as the main fit. These results are consistent with removing a $\sim\!1~\mathrm{keV}$ component from these outskirts regions, i.e., in the case of multi-structure components, leaving a hotter component for the A3391, and a $\sim\!1~\mathrm{keV}$ component for A3395, because the temperatures found for A3395 and BGSFwoLSC are of this order. The unexpectedly high lower temperature limit for the A3391-N BGNF makes this model seem rather unlikely.
\par
To check whether the filament emission is of Galactic origin, we performed the following three tests for the model in the filament regions; first to fix the redshift of the filament component to 0, second to limit the redshift within $\pm15\%$ of the average of the cluster redshift, third to limit the redshift between 0 and 1. The normalizations, temperatures, and metallicities of the first and second cases are in good agreement with the main fit. The third case yields unexpectedly high redshifts, that is $z>0.97$ and $z=0.19$ for NF and SF, respectively. This would result in an improbably long physical extent of these filaments, namely at least 13.7 Mpc for NF and 8.3 Mpc for SF. As redshift estimation strongly depends on emission lines, these high redshift results could be due to statistical fluctuations. Additionally, we performed a fit where the redshift is fixed at 0 and metallicity to a solar value (while in the above tests, the metallicity is left to vary). The test could only be done to the SF and SFwoLSC, while the NF fit did not converge. This test results in lower temperature limits of $4.6~\mathrm{keV}$ ($4.4~\mathrm{keV}$) for the SF (SFwoLSC). These values are highly unlikely for a region outside $R_{200}$ and are also unlikely temperatures for any extended emission coming from the Galaxy. Thus, this fit cannot physically describe the fitted spectra.
\par
Lastly, we swapped the sky background regions used for the fits in the northern and southern regions. As presented in Table~\ref{tab:sky_BG}, we see in general that the normalizations of the soft components (LHB and MWH) of the CXB-S are higher than those of CXB-N. This is also reflected in the results of this test. The normalization of the NF using CXB-S as the sky background is lower, and the temperature gets higher. On the contrary, as a result of underestimating the soft sky background components, the normalization yielded for the SF (SFwoLSC) using CXB-N as the sky background is higher and the resulting overdensity values indicate that the gas is HM, which is unforeseen for this region. This test illustrates that our default definition of the background regions seems appropriate.
\par
Description of the fitting methods shown in Table~\ref{tab:spectro_tests}: BGNF: Using Northern Filament boxes as sky background. 2T: Two-temperature components. $z=0$: Redshift is fixed to 0. $z_{\mathrm{free, range \pm15\%}}$: Redshift is left to vary in a range of $\pm15\%$ of the starting value. $z_{\mathrm{free, range0-1}}$: Redshift is left to vary between 0 and 1. CXB-S: Using CXB-S circle as background (Fig.~\ref{fig:spectro_regions} and \ref{fig:NH}). BGSFwoLSC: Using Southern Filament boxes (LSC excluded) as sky background. CXB-N: Using the CXB-N circle as background (Fig.~\ref{fig:spectro_regions} and \ref{fig:NH}). $z=0,~Z=1$: Redshift is fixed to 0 and metallicity to $1Z_\odot$.

\begin{table}[!h]
    \centering
    \caption{eROSITA spectral analysis results from additional tests, the derived electron density ($n_\mathrm{e}$), and the filament gas overdensity ($\delta_b$). }
    \resizebox{\textwidth}{!}{\begin{tabular}{c c c c c c c c}
    \hline
    \hline
\multirow{2}{*}{Regions} & \multirow{2}{*}{Fitting} & $norm$ & $k_\mathrm{B}T$ & $Z$ & $n_\mathrm{e}$ &  \multirow{2}{*}{$\delta_b$} & \multirow{2}{*}{stat/dof}\\
 & & $[10^{-6}$ cm$^{-5}/$arcmin$^2]$ & $[$keV$]$ & $[Z_{\odot}]$ & $[10^{-5}$ cm$^{-3}]$ & \\[5pt]
\hline \\[-1.7ex]
\hline \\[-1.7ex]
\multicolumn{8}{c}{\large{NORTH}}\\
\hline \\[-1.7ex]
\hline \\[-1.7ex]
A3391-N & BGNF & $1.18_{-0.21}^{+0.35}$ & $>4.52$ & $>0.09$ & & & 7989.7/9087 \\[5pt]
\hline \\[-1.7ex]
NC & BGNF & $1.78_{-0.84}^{+0.79}$ & $0.98_{-0.19}^{+0.29}$ & $0.14_{-0.09}^{+0.14}$ & & & 5682.5/6104\\[5pt]
\hline \\[-1.7ex]
\multirow{2}{*}{Box1+2} & \multirow{2}{*}{2T} & $1.29_{-1.16}^{+3.39}$ & $0.14_{-0.02}^{+0.03}$ & $0.22_{-0.17}^{+0.62}$ & $5.24_{-3.56}^{+4.73}$ & $199_{-135}^{+179}$ & \multirow{2}{*}{10938.4/10740} \\[5pt]
& & $1.05_{-0.21}^{+0.27}$ & $1.28_{-0.23}^{+0.69}$ & $<0.12$ & $4.73_{-0.49}^{+0.57}$ & $179_{-19}^{+21}$ & \\[5pt]
\hline \\[-1.7ex]
Box1+2 & $z=0$ & $1.51_{-0.16}^{+0.17}$ & $0.89_{-0.11}^{+0.16}$ & $<0.01$ &  &  & 10874.7/10743 \\[5pt]
\hline \\[-1.7ex]
Box1+2 & $z_{\mathrm{free, range \pm15\%}}=0.061$ & $1.67_{-0.21}^{+0.2}$ & $0.97_{-0.17}^{+0.16}$ & $<0.01$ &  &  & 10854.9/10742  \\[5pt]
\hline \\[-1.7ex]
Box1+2 & $z_{\mathrm{free, range 0-1}}>0.967$ & $4.06_{-0.8}^{+1.04}$ & $1.56_{-0.19}^{+0.14}$ & $0.25_{-0.14}^{+0.23}$ & & & 10843.0/10742 \\[5pt]
\hline \\[-1.7ex]
Box1+2 & CXB-S & $0.47_{-0.11}^{+0.14}$ & $>1.88$ & $>0.41$ & $3.17_{-0.39}^{+0.43}$ & $120_{-15}^{+16}$ & 10872.3/10790 \\[5pt]
\hline \\[-1.7ex]
\hline \\[-1.7ex]
\multicolumn{8}{c}{\large{SOUTH}}\\
\hline \\[-1.7ex]
\hline \\[-1.7ex]
A3395 & BGSFwoLSC & $0.22_{-0.05}^{+0.14}$ & $1.68_{-0.16}^{+0.25}$ & $>0.97$ & & & 10320.1/10701\\[5pt]
\hline \\[-1.7ex]
Box1+2 & $z=0$ & $0.71_{-0.12}^{+0.13}$ & $0.95_{-0.08}^{+0.07}$ & $0.07_{-0.03}^{+0.04}$ & & & 12821.9/11368 \\[5pt]
\hline \\[-1.7ex]
Box1+2 & $z_{\mathrm{free, range \pm15\%}}>0.053$ & $0.74_{-0.12}^{+0.13}$ & $1.09_{-0.04}^{+0.11}$ & $0.1_{-0.04}^{+0.06}$ & & & 12755.8/11367\\[5pt]
\hline \\[-1.7ex]
Box1+2 & $z_{\mathrm{free, range 0-1}}=0.193_{-0.020}^{+0.006}$ & $0.62_{-0.11}^{+0.1}$ & $1.91_{-0.23}^{+0.23}$ & $0.68_{-0.40}^{+0.44}$ & & & 12753.8/11367\\[5pt]
\hline \\[-1.7ex]
Box1+2 & $z=0,~Z=1$ & $0.51_{-0.04}^{+0.04}$ & $>4.61$ & $1.0$ &  & & 13087.4/11369 \\[5pt]
\hline \\[-1.7ex]
Box1+2 & CXB-N & $2.80_{-0.33}^{+0.32}$ & $0.51_{-0.04}^{+0.06}$ & $0.01_{-0.01}^{+0.01}$ & $8.52_{-0.52}^{+0.48}$ & $324_{-20}^{+18}$ & 12882.1/11321\\[5pt]
\hline \\[-1.7ex]
Box1+2, woLSC & $z=0$ & $0.40_{-0.12}^{+0.11}$ & $1.05_{-0.15}^{+0.19}$ & $0.06_{-0.05}^{+0.08}$ & & & 12675.4/11363\\[5pt]
\hline \\[-1.7ex]
Box1+2, woLSC & $z_{\mathrm{free, range \pm15\%}}=0.060$ &  $0.39_{-0.12}^{+0.13}$ & $1.19_{-0.15}^{+0.18}$ & $0.12_{-0.08}^{+0.15}$ & & & 12619.9/11362\\[5pt]
\hline \\[-1.7ex]
Box1+2, woLSC & $z=0, Z=1$ &  $0.29_{-0.04}^{+0.04}$ & $>4.38$ & $1.0$ & & & 12889.0/11364\\[5pt]
\hline \\[-1.7ex]
Box1+2, woLSC & CXB-N &  $2.79_{-0.44}^{+0.44}$ & $0.38_{-0.05}^{+0.06}$ & $0.01_{-0.01}^{+0.02}$ & $8.19_{-0.67}^{+0.62}$ & $311_{-25}^{+24}$ & 12712.7/11316\\[5pt]
\hline
\hline
    \end{tabular}}
    \label{tab:spectro_tests}
\end{table}

\end{appendix}

\end{document}